\documentclass[%
 reprint,
 nofootinbib,
 nobibnotes,
 amsmath,amssymb,
 aps,
 pra,
]{revtex4-2}

\usepackage{graphicx}
\usepackage{dcolumn}
\usepackage{bm}
\usepackage[hidelinks]{hyperref}
\usepackage{gensymb}

\newcommand{\ket}[1]{|{#1}\rangle}
\newcommand{\bra}[1]{\langle{#1}|}

\begin{document}

\preprint{APS/123-QED}

\title{A Laser-Atom Interaction Simulator derived from Quantum Electrodynamics}

\author{Manish Patel}
\author{Matthew Harvey}
\author{Andrew James Murray}
 \email{andrew.murray@manchester.ac.uk}
\affiliation{
 Photon Science Institute, Department of Physics and Astronomy\\
 University of Manchester, 
 Manchester, M13 9PL, UK
}

\date{May 10, 2022}
             
\begin{abstract}
    A laser-atom interaction simulator derived from quantum electrodynamics (LASED) is presented, which has been developed in the python programming language. LASED allows a user to calculate the time evolution of a laser-excited atomic system. The model allows for any laser polarization, a Gaussian laser beam profile, a rotation of the reference frame chosen to define the states, and an averaging over the Doppler profile of an atomic beam. Examples of simulations using LASED are presented for excitation of calcium from the 4$^1S_0$ state to the 4$^1P_1$ state, for excitation from the helium 3$^1D_2$ state excited by electron impact to the 10$^1P_1$ state, and for laser excitation of caesium via the $D_2$ line. 
\end{abstract}

\maketitle

\section{Introduction}
Laser-excitation of atoms is an essential physical process used in many experiments including spectroscopy \cite{neugart2017collinear, studer2019atomic}, trapping of atoms \cite{Raab1987,harvey2008cold,cortinas2020laser, urvoy2019direct}, collision physics \cite{HertelStoll,farrell1988quantum,Macgillivray_Standage1988,murraystepwisePRL,Stepwise_1990,
farrell1991quantum,murraystepwiseJModOpt,Stepwise_1992_expt,murray_radiation_trapping,masters1996multipole,
murray2003low,superelasticBfieldPRL,murray2008theoretical,superelasticBfieldJPB,e2eMgPRL1,e2eMgPRL2} and atomic interferometry \cite{rudolph2020large}. To design experiments such as these, modelling of the dynamics of the laser-atom interaction with time is often required. The semi-classical approach to solving the equation of motions of these systems has been used extensively, where the atom is quantized and the field is treated classically \cite{mcClelland,robertson2021arc, stenholm1986semiclassical, milovsevic2017semiclassical}. In these models the relaxation terms are added phenomenologically. By contrast, models that treat both the atom and field quantum mechanically can describe the relaxation terms more rigorously, particularly when the system is complex with many substates involved in the interaction \cite{farrell1988quantum}. Deriving the equations of motion (often called the optical Bloch equations) by hand and solving the laser-atom interaction is time-intensive, complex, and is prone to mistakes. As an example, for transitions with hyperfine structure such as excitation of the Cs 6$^2S_{1/2}$ state to the 6$^2P_{3/2}$ state, a total of 48 individual substates are involved in the interaction. There are hence 2304 coupled differential equations that must be generated and solved simultaneously to fully characterise the dynamics of the system. A computational method of systematically generating and solving these equations is hence advantageous, so that the time evolution of the populations, optical coherences and the atomic coherences can be obtained. 

This paper presents an open-source python package that solves this problem: a laser-atom interaction simulator derived from quantum electrodynamics (LASED). LASED allows a user to automatically set up a laser-atom system and generate all the equations of motion for that system, which can be printed out in LaTeX. The package then solves the dynamics of the system over a given time, outputting the evolution of all lower and upper state populations, their atomic coherences and the optical coherence terms that couple the states together. LASED can also model the system using laser beams that have different polarizations. It can model a rotation of the frame of reference of the system both prior to the interaction, as well as after the laser interaction has occurred. This rotation technique can simplify the calculation, thereby reducing the time required for generating an output. LASED can further include integration over the Gaussian profile of the laser beam (assuming a TEM$_{00}$ beam) and also allows integration over the Doppler profile of an atomic beam, should this be required for the experiment that is being modelled. The angular `shape' of the electron charge cloud for both excited and lower atomic states can also be modelled and plotted as the system evolves over time.

Other laser-atom interaction simulators exist that are used to describe different processes. The simulator described in \cite{eckel2022pylcp} has been developed to model atoms that are laser-cooled in a magneto-optical trap, and includes the effect of the trapping magnetic field as well as the laser field. LASED has been developed in a similar way, however it can also describe different experiments such as scattering experiments that combine laser interactions with electron collisions and that use an atomic beam \cite{HertelStoll,farrell1988quantum,Macgillivray_Standage1988,murraystepwisePRL,Stepwise_1990,
farrell1991quantum,murraystepwiseJModOpt,Stepwise_1992_expt,murray_radiation_trapping,masters1996multipole,
murray2003low,superelasticBfieldPRL,murray2008theoretical,superelasticBfieldJPB,e2eMgPRL1,e2eMgPRL2}. LASED is designed to be easy-to-use and has comprehensive online documentation to aid users in creating the required laser-atom system they wish to model. This documentation also demonstrates how to run the simulations by solving examples of the differential equations automatically generated by LASED \cite{LASEDreadthedocs}. Details on how to install LASED can be found in this documentation and in appendix \ref{appendix:installation}.

In this paper the derivation of the general equations of motion which are adopted in LASED are briefly presented in section \ref{section:eq_of_motion}. In section \ref{section:time_evol} the computational method for generating the coupled differential equations to solve the time evolution of the laser-atom system is discussed. This section also shows how averaging over the Gaussian and Doppler profiles is approximated, and it details how the reference frame is rotated. The method used to model a general polarization state of the laser is also described. 

Section \ref{section:examples} demonstrates the outcome from the model for three selected targets. In the first example excitation of calcium from the ground 4$S$ state to the 4$P$ state is discussed, since this is one of the simplest systems that can be solved. These results are presented for both linear and elliptic excitation, and includes integration over both Doppler and Gaussian profiles. An example of the technique of rotating the frame of reference is then discussed, with the calcium target again being used. In the second example, laser excitation of helium initially excited by electron impact to a $D$-state is presented. This is a considerably more complex problem to solve, since the lower state is then in a coherent superposition of substates due to the collision. Both the populations and atomic coherences are hence non-zero prior to laser excitation and these must be included as initial conditions. Finally, a discussion of laser excitation of the Cs atom from the ground state via the $D_2$ transition is presented for circular excitation, as would be used in a Magneto Optical Trap (MOT).

\section{General Equations of Motion} \label{section:eq_of_motion}

The general equations of motion using the QED approach for continuous wave laser irradiation of atoms have been detailed in \cite{farrell1988quantum,farrell1991quantum,Macgillivray_Standage1988,murray2008theoretical}, and so only a brief overview is presented here. The equations are derived using the Heisenberg formulation, where the operators are chosen to evolve in time. The Hamiltonian of the laser-atom system is hence given by
\begin{equation}
    H = H_{\textrm{Atom}}+H_{\textrm{Field}}+H_{\textrm{Int}}
\end{equation}
where $H_{\textrm{Atom}}$ describes the atom evolving freely over time, so that
\begin{equation} \label{eq:Hatom}
    H_{\textrm{Atom}} = \sum_{i}\hbar\omega_i\ket{i}\bra{i}
\end{equation}
where $\hbar\omega_i$ is the energy of the $i$'th level. The Hamiltonian for the field is represented quantum mechanically by
\begin{equation}
    H_{\textrm{Field}} = \sum_{q}\hbar\omega_qa_{q}^{\dagger}a_{q}
\end{equation}
where $a_{q}^{\dagger}$,$a_{q}$ are the creation and annihilation operators for the mode $q$ of the field, with $q$ representing both the wave vector and its polarization. The interaction Hamiltonian \cite{loudon2000quantum} is expressed in normal ordering \cite{ackerhalt1974quantum} as
\begin{equation} \label{eq:Hint}
H_{\textrm{Int}} = \hbar\sum_{e'g'q'}g^{q'}_{e'g'}\hat{\sigma}_{e'g'}a_{q'}e^{ik_{q'}z} + g^{q'*}_{e'g'}a^{\dagger}_{q'}e^{-ik_{q'}z}\hat{\sigma}_{g'e'} 
\end{equation}
where 
\begin{equation} \label{eq:couplingcoeff}
   g^{q'}_{e'g'} = i\sqrt{\frac{\omega_{q'}}{2\epsilon_0\hbar{V}}}\hat{\textbf{e}}_{q'}.\textbf{D}_{e'g'}
\end{equation}
is a coupling coefficient between the mode of the laser field and the lower atomic state $\ket{g'}$ and upper state $\ket{e'}$. $V$ is the mode volume, $\textbf{D}_{e'g'}$ is the dipole moment, and $\hat{\textbf{e}}_{q'}$ is the polarization unit vector. The atomic operators are defined by the atomic states so that
\begin{equation}
    \hat{\sigma}_{eg} = \ket{e}\bra{g}.
\end{equation}
where $\ket{g}$ represents the manifold of all lower states of the system and $\ket{e}$ represents the manifold of upper states that are coupled to $\ket{g}$ by the laser. The atomic operators evolve over time using the Liouville equation \cite{kubo1963stochastic}
\begin{align} \label{eq:liouville}
     \frac{d\hat{\sigma}_{eg}}{dt} & = -\frac{i}{\hbar}[\hat{\sigma}_{eg}, H] \nonumber \\ 
    & = -\frac{i}{\hbar}[\hat{\sigma}_{eg}, H_{Atom}] -\frac{i}{\hbar}[\hat{\sigma}_{eg}, H_{Int}].
\end{align}
$H_{\textrm{Field}}$ does not contribute here as it commutes with the atomic operator. The first term in equation \ref{eq:liouville} can be simplified using the ortho-normality relations $\bra{e}e\rangle = \bra{g}g\rangle = 1$ and $\bra{e}g\rangle = \bra{g}e\rangle = 0$ so that
\begin{equation} \label{eq:1stterm}
    -\frac{i}{\hbar}[\hat{\sigma}_{eg}, H_{\textrm{Atom}}] = -i(\omega_g - \omega_e)\ket{e}\bra{g}
\end{equation}
The second term in equation \ref{eq:liouville} can be expanded using equation \ref{eq:Hint} so that
\begin{align} \label{eq:Hint2ndtermexpanded}
   -\frac{i}{\hbar}[\hat{\sigma}_{eg}, H_{\textrm{Int}}] =  & \nonumber -i\sum_{e'q'}g^{q'*}_{e'g}a_{q'}(t)^{\dagger}e^{-ik_{q'}z}\hat{\sigma_{ee'}} \\ + & i\sum_{g'q'}g^{q'*}_{eg'}a_q^{\dagger}e^{-ik_qz}\hat{\sigma}_{gg'}
\end{align}
As the time evolution of the annihilation and creation operators depends on the field coupling to the atomic states, an explicit function in time is required. For $a_{q}^{\dagger}$ this is given by
\begin{align} \label{eq:annihilationoperator}
    a^{\dagger}_{q'}(t) = & a^{\dagger}_{q'}(0)e^{i\omega_{q'}t} \\ & \nonumber +i\sum_{e''g''}g^{q'}_{e''g''}e^{ik_{q'}z}\int^t_0\hat{\sigma}_{e''g''}(t')e^{i\omega_{q'}(t-t')}dt
\end{align}
with $a_{q}$ given by the complex conjugate of this equation. In equation \ref{eq:annihilationoperator} the atomic operator can be removed from the integral using the  Harmonic approximation \cite{whitley1976double}. When combined with equation \ref{eq:Hint2ndtermexpanded}, this then yields
\begin{align} \label{eq:Hint2ndtermfinal}
   -\frac{i}{\hbar}[\hat{\sigma}_{eg}, H_{\textrm{Int}}] =   & -i\sum_{q'e'}g^{q'*}_{e'g}a_{q'}^{\dagger}(0)e^{i({\omega_{q'}t-k_{q'}}z)}\hat\sigma_{ee'} \nonumber \\
    & + \sum_{q'e'e''g''}\Big(g^{q'*}_{e'g}g^{q'}_{e''g''}\hat{\sigma}_{e''g''} \nonumber \\ & \times \int^t_0e^{i(\omega_{q'}-\omega_{e''}+\omega_{g''}(t-t'))}dt'\Big) \hat\sigma_{ee'} \nonumber \\
    & + i\sum_{q'g'}g^{q'}_{eg'}a^{\dagger}_{q'}(0)e^{i\omega_{q'}t-k_{q'z}}\hat\sigma_{g'g} \nonumber \\
   & - \sum_{q'g'e''g''} \Big(g^{q'*}_{eg'}g^{q'}_{e''g''}{\hat\sigma_{e''g''}} \nonumber \\
   & \times \int^t_0e^{i(\omega_{q'}-\omega_{e''}+\omega_{g''}(t-t'))}dt'\Big)\hat\sigma_{g'g}.
\end{align}
When the laser frequency $\omega_{q'}$ is close to the transition frequency $\omega_{e''}- \omega_{g''}$ and for time periods much larger than the inverse of the oscillation frequency, the integrals in equation \ref{eq:Hint2ndtermfinal} can be approximated to $\delta$ functions \cite{ackerhalt1974quantum}. The time evolution of the atomic operator can hence be written as 
\begin{align} \label{eq:evolutionatomicoperators}
    \frac{d\hat{\sigma}_{eg}}{dt} = & -i(\omega_g - \omega_e){\hat\sigma_{eg}} \nonumber \\
    & -i\sum_{q'e'}g^{q'}_{e'g}a^{\dagger}_{q'}(0)e^{i(\omega_{q'}t-k_{q'}z)} \hat{\sigma}_{ee'} \nonumber \\
    & + i\sum_{q'g}g^{q'*}_{eg'}a^{\dagger}_{q'}(0)e^{i(\omega_{q't-k_{q'}z})}\hat\sigma_{g'g} \nonumber \\
    & - \sum_{q'g'e'}g^{q'*}_{eg'}g^{q'}_{e'g'}\hat{\sigma}_{e'g}\pi\delta(\omega_{q'}-\omega_{e'}+\omega_{g'}).
\end{align}
Equation \ref{eq:evolutionatomicoperators} contains rapidly oscillating terms at the frequency of the driving radiation. In many experiments these cannot be measured, and so the Rotating Wave Approximation (RWA) \cite{whitley1976double} is adopted. For a single mode continuous wave laser beam driving the transition, the RWA transforms the atomic operators into slowly varying operators $\hat{\chi}_{eg}$, by setting
\begin{align}
    & \hat{\sigma}_{eg} = \hat{\chi}_{eg}e^{i(\omega_Lt-k_Lz)} \label{eq:chi_eg_transform} \\
    & \hat{\sigma_{gg'}} = \hat{\chi}_{gg'} \label{eq:chi_gg_transform}\\
    & \hat{\sigma_{ee'}} = \hat{\chi}_{ee'} \label{eq:chi_ee_transform},
\end{align}
where $\omega_L$ is the laser frequency and $k_L$ is the associated wave vector. Equation \ref{eq:evolutionatomicoperators} is hence transformed to slowly varying operators using equation \ref{eq:chi_eg_transform}. Expectation values are then taken, so that 
\begin{align} \label{eq:chi_eg}
    \langle\dot{\hat{\chi}}_{eg}\rangle = & -i(\omega_L-k_L\dot{z}-\omega_{eg})\langle\hat{\chi}_{eg}\rangle \nonumber \\
    & -i\sum_{Le'}g^{L*}_{e'g}\langle{a_L^{\dagger}(0)}\rangle \langle\hat{\chi}_{ee'}\rangle \nonumber \\
    & + i\sum_{Lg'}g^{L*}_{eg'}\langle a_L^{\dagger}(0)\rangle \langle\hat{\chi}_{g'g}\rangle \nonumber \\
    & - \sum_{qg'e'}g^{q*}_{eg'}g^q_{e'g'}\langle\hat{\chi}_{e'g}\rangle\pi\delta(\omega_L-\omega_{e'}+\omega_{g'}).
\end{align}

The slowly varying operators are directly related to the density matrix elements $\rho_{eg}$ that are commonly used to describe the populations and coherences of an atomic system, since
\begin{equation} \label{eq:chi_hat_rho_relation}
    \langle\hat{\chi}_{eg}\rangle = \bra{\psi}e\rangle\bra{g}\psi\rangle = (\bra{e}\psi\rangle\bra{\psi}g\rangle)^* = (\rho_{eg})^{*} = \rho_{ge}.
\end{equation}
The half-Rabi frequency is input to equation \ref{eq:chi_eg} using the relation \cite{farrell1988quantum}
\begin{equation} \label{eq:halfRabicreation}
    \Omega^L_{eg} = g^{L*}_{eg}\langle\hat{a}_{L}^{\dagger}(0)\rangle.
\end{equation}
This is set to be real by an appropriate choice of phase \cite{farrell1988quantum}. 

The time evolution of $\langle\dot{\hat{\chi}}_{gg''}\rangle$ and $\langle\dot{\hat{\chi}}_{ee''}\rangle$ can be derived in an identical way to that of $\langle\dot{\hat{\chi}}_{eg}\rangle$ in equation \ref{eq:chi_eg}. These can then be written in the density matrix formalism using equation \ref{eq:chi_hat_rho_relation}. The general equations of motion for the populations, optical and atomic coherences as used in LASED are then given by:
\begin{align}
    \dot{\rho}_{gg''} = & -i\Delta_{gg''}\rho_{gg''} + i\sum_{Le}(\Omega^L_{eg''}\rho_{ge} - \Omega^L_{eg}\rho_{eg''}) \nonumber \\
    & + \sum_{qe'e''}g^q_{e'g''}g^{q*}_{e''g}\pi\delta(\omega_q-\Delta_{e''g})\rho_{e''e'} \nonumber \\
    &  + \sum_{qe'e''}g^q_{e''g''}g^{q*}_{e'g}\pi\delta(\omega_q-\Delta_{e''g''})\rho_{e'e''} \label{eq:rho_gg}
\end{align}
\begin{align}
    \dot{\rho}_{ee''} = & -i\Delta_{ee''}\rho_{ee''} + i\sum_{Lg}(\Omega^L_{e''g}\rho_{eg} - \Omega^L_{eg}\rho_{ge''}) \nonumber \\
    & -\sum_{qg'e'}g^q_{eg'}g^{q*}_{e'g'}\pi\delta(\omega_q - \Delta_{e'g'})\rho_{e'e''}) \nonumber \\
    & - \sum_{qg'e'}g_{e'g'}^{q}g_{e''g'}^{q*}\pi\delta(\omega_q - \Delta_{e'g'})\rho_{ee'} \label{eq:rho_ee}
\end{align}
\begin{align}
    \dot{\rho}_{ge} = & -i\Delta_{L,eg}\rho_{ge} - i\sum_{Le'}\Omega_{e'g}^{L}\rho_{e'e} + i\sum_{Lg'}\Omega_{eg'}^L\rho_{gg'}  \nonumber \\
    & - \sum_{qg'e'}g_{e'g'}^{q}g_{eg'}^{q*}\pi\delta(\omega_q - \Delta_{e'g'})\rho_{ge'} \label{eq:rho_ge}
\end{align}
with $\dot{\rho}_{eg}$ given by the complex conjugate of equation \ref{eq:rho_ge}.  

The population equations for the lower and upper states are derived by setting $g = g'$ and $e=e'$ respectively in equations \ref{eq:rho_gg} and \ref{eq:rho_ee}. The atomic coherence equations (which describe the phase relationship between sub-states in each manifold) are formulated by setting $g \neq g'$ and $e \neq e'$ in these equations. The optical and non-optical terms between the lower and upper states manifolds are described by equation \ref{eq:rho_ge}. This equation hence is used to derive both the optical coherence terms generated directly by the laser, as well as the atomic coherence terms generated between upper and lower states.  The term
\begin{equation} \label{eq:doppler_detuning}
    \Delta_{L,eg} = \omega_L - \frac{2\pi v_z}{\lambda_L} + \omega_e - \omega_g
\end{equation}
is the detuning from resonance, where $v_z$ is the velocity component of the atoms in the direction of the laser beam (which gives rise to the Doppler shift). $\lambda_L$ is the wavelength of the laser mode $L$. The  term $\Delta_{eg} = \omega_e - \omega_g$, whereas $\Delta_{gg'}= \omega_g - \omega_{g'}$ and $\Delta_{ee'}= \omega_e - \omega_{e'}$. The half-Rabi frequency can be calculated in rad/s using the expression 
\begin{equation} \label{eq:rabifreq}
    \Omega_{eg}^q = C_{eg}^q\Omega = C_{eg}^q\sqrt{\frac{3\lambda_L^3I_L}{8{\pi}hc\tau}}
\end{equation}
where $\tau$ is the lifetime of the transition, $I_L$ is the laser intensity, and $C^q_{eg}$ is a coupling coefficient given by \cite{farrell1995consistency}
\begin{align}
    C^q_{eg} = & (-1)^{\frac{q(1+q)}{2}+F'+F+J'+J+I'+L'+S'-m_F'+1} \nonumber \\
   & \times \sqrt{(2F'+1)(2F+1)(2J'+1)(2J+1)(2L'+1)} \nonumber \\
   & \times
\begin{pmatrix} 
F' & 1 & F \\
-m_F' & q & m_{F} 
\end{pmatrix}
\begin{Bmatrix}
J' & F' & I' \\
F & J & 1
\end{Bmatrix}
\begin{Bmatrix}
L' & J' & S'  \\
J & L  & 1 
\end{Bmatrix}
\end{align}
where $L, S, J, I$, and $F$ are the quantum numbers describing the lower states $\ket{g}$ and their primed equivalents are the quantum numbers describing the upper states $\ket{e}$. $q$ is set to be either +1, 0, or -1 for the laser polarization being  right-hand circularly polarized (RHC), linearly polarized, or left-hand circularly polarized (LHC) respectively. 

The triple summations in equations \ref{eq:rho_gg}, \ref{eq:rho_ee}, and \ref{eq:rho_ge} describe spontaneous emission. They produce decay of the atomic excitation even if there is no driving laser field. These terms can be calculated by relating them to the generalized decay rate given in \cite{farrell1988quantum}
\begin{align}
    \Gamma_{ege'g'} = & \sum_q[g^q_{eg'}g^{q*}_{e'g}\pi\delta(\omega_q-\Delta_{e'g'}) \nonumber \\
    & + g^q_{e'g'}g^{q*}_{eg}\pi\delta(\omega_q-\Delta_{eg})] \label{eq:Gamma_egeg}.
\end{align}
Equation \ref{eq:Gamma_egeg} is then used to derive the decay rate between an excited substate $\ket{e}$ to a lower substate $\ket{g}$ with
\begin{equation} \label{eq:Gamma_eg}
    \Gamma_{eg} = \Gamma_{egeg} = 2\sum_{q}|g^q_{eg}|^2\pi\delta(\omega_q-\Delta_{eg}) 
\end{equation}
The total decay rate of state $\ket{e}$ is then given by
\begin{equation} \label{eq:Gamma_e}
    \Gamma_e = \sum_g\Gamma_{eg}.
\end{equation}
The transition probability for spontaneous emission is proportional to the square of the dipole matrix element and so the decay constants $\Gamma_{eg}$ can be calculated using
\begin{equation}
    \Gamma_{eg} = \frac{(\Omega_{eg})^2}{\sum_{g'}(\Omega_{eg'})^2}\Gamma_e = \frac{|C^q_{eg}|^2}{\tau\sum_{g'}|C^q_{eg'}|^2}
\end{equation}
where the summation in the denominator is over all coupled ground states and $\tau$ is the lifetime of the excited state. Here, $q$ is the required polarization for the decay from $\ket{e}$ to $\ket{g}$.

The equations presented above are then used to compute the time evolution of the laser-atom system, as detailed in the next section.

\begin{figure}[b]
    \centering
    \includegraphics[scale=0.45]{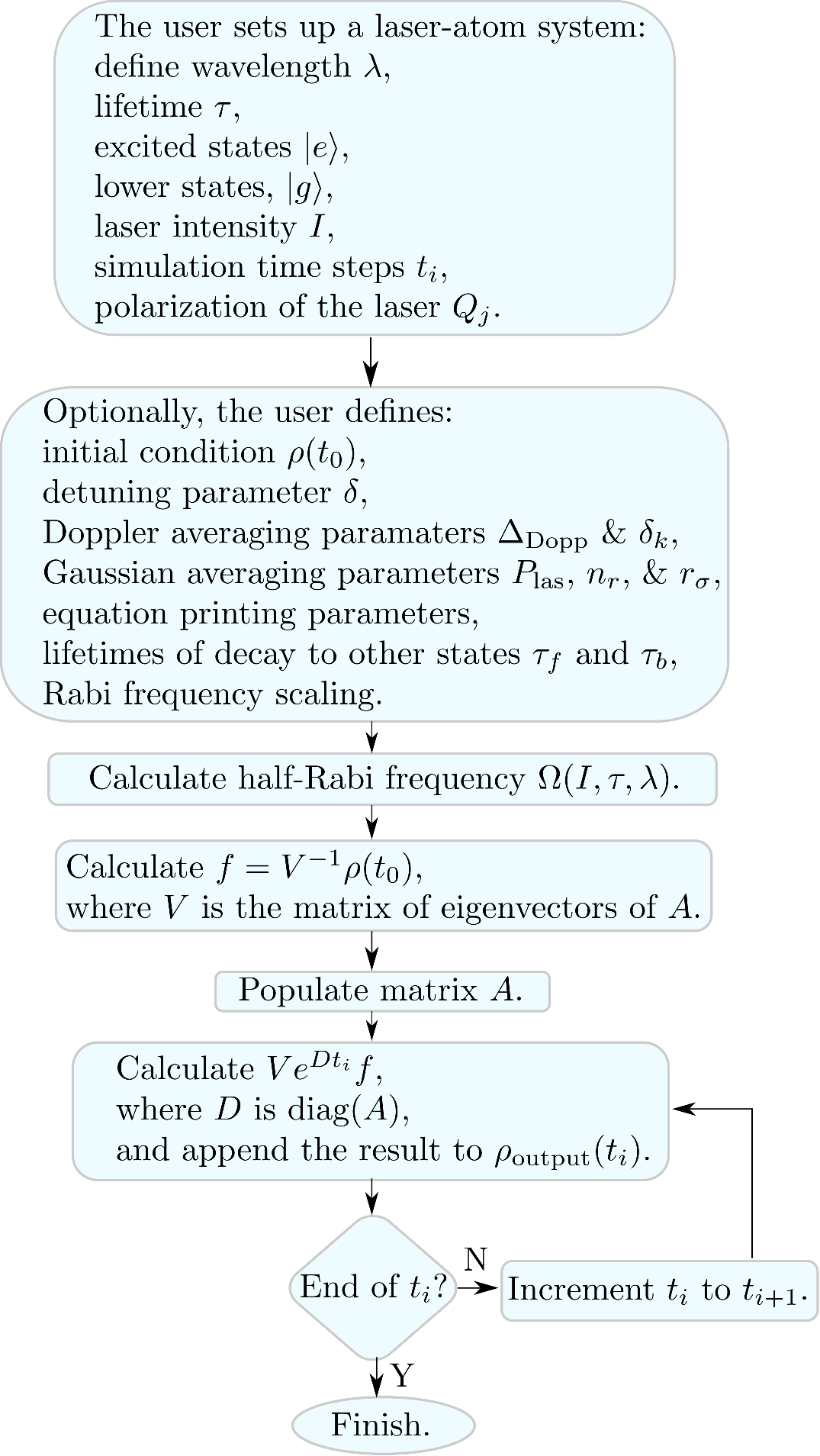}
    \caption{The algorithm used in LASED to compute the time evolution of a laser-atom system. For details, see the text in section \ref{section:time_evol}.}
    \label{fig:algorithm}
\end{figure}

\section{Time Evolution Calculations} \label{section:time_evol}
The process of setting up a laser-atom system and finding the solutions to the equations of motion is outlined in Fig. \ref{fig:algorithm}. LASED enables a user to define an atomic system by creating the states and substates of the atom that are coupled by the laser. Their relative energy separation, angular momenta, and the projection of the total angular momentum associated with each state are also input to the model as initial parameters. The substates are labelled as either an upper state $\ket{e}$ or a lower state $\ket{g}$. The resonant transition laser wavelength between upper and lower states is defined as $\lambda$. The user then enters the laser parameters by defining the laser polarization $Q$ and the intensity $I$. This sets up the initial laser-atom system to be solved. The time steps $t_i$ over which the simulation is run is also defined before the system evolves. At the initial time step $t = t_0$ the laser is turned on.

To solve the user-defined laser-atom system, equations \ref{eq:rho_gg}, \ref{eq:rho_ee} and \ref{eq:rho_ge} are used to automatically generate the complete set of coupled differential equations which are solved numerically. LASED uses a matrix method by writing the equations in the form
\begin{equation} \label{eq:rho_matrix}
    \dot{\rho} = A\rho(t)
\end{equation}
where $\rho$ is a column vector containing all the populations and coherences defined within the density matrix for the coupled system. The density matrix has $n^2$ elements, where $n$ is the number of substates in the system. $\rho$ is hence a column vector of $n^2$ elements. $A$ is an $n^2\times{n^2}$ coupling matrix that contains all of the coefficients of the interaction. This includes all half-Rabi frequencies generated from equation \ref{eq:rabifreq}, the detuning terms and all decay constants. The matrix $A$ can become very large, and so to reduce computation time equation \ref{eq:rho_matrix} is solved by diagonalising the matrix and calculating the eigenvectors and eigenvalues. The solution of $\rho$ using this technique then gives
\begin{equation} \label{eq:rho_matrix_exp}
\rho(t) = Ve^{Dt}V^{-1}\rho(t_0)
\end{equation}
where $A = VDV^{-1}$. $D$ is the diagonalised form of $A$ that contains the complex eigenvalues and $V$ is the matrix of eigenvectors of $A$. All real terms in the eigenvalues generated by the calculation must be negative for the
solutions to converge. The initial condition $\rho(t_0)$ can be defined by the user when setting up the laser-atom system. If the initial conditions are not defined by the user, it is assumed that all lower substates have equal populations and that all atomic coherences are identically zero prior to the laser being turned on (as would occur if the atomic beam was generated from an oven or from a gas jet).

Using this matrix method requires the populations and coherences to be in a strict order within $\rho$. The user hence has to define the substate with a number that labels it, e.g. $\ket{1}$, $\ket{2}$, and $\ket{3}$ for a lower $P$-state with ${m = -1, 0, +1}$ respectively. The convention used throughout LASED is that the first element in the vector $\rho$ is the element $\rho_{11}$, which is the lower state population with the lowest projection of angular momentum $-m_F$. The labelling continues until the excited substate population with the largest projection of total angular momentum $+m_F$ is set to be the $n^{th}$ substate. The vector would hence have the form: [$\rho_{11}$ $\rho_{12}$ ... $\rho_{1n}$, $\rho_{21}$ $\rho_{22}$ ... $\rho_{2n}$, $\rho_{n1}$ $\rho_{n2}$ ... $\rho_{nn}$]. The matrix $A$ is then populated using the coefficients generated using equations \ref{eq:rho_gg} to \ref{eq:rho_ge}. These can be modified to give a set of equations with computable variables:
\begin{align}
    \dot{\rho}_{gg''} = & -(i\Delta_{gg''} + \frac{1}{\tau_b})\rho_{gg''} \nonumber \\ 
    & + i\Omega\sum_{qe}(C^q_{eg''}\rho_{ge} - C^q_{eg}\rho_{eg''}) \nonumber \\
    & + \frac{1}{2\tau}\sum_{qe'e''}\gamma^q_{e'e''}\rho_{e''e'} + \gamma^q_{e''e'}\rho_{e'e''} \label{eq:rho_gg_lased}
\end{align}
\begin{align}
    \dot{\rho}_{ee''} = & -(i\Delta_{ee''}+\frac{1}{\tau}+\frac{1}{\tau_f})\rho_{ee''} \nonumber \\
    & + i\Omega\sum_{qg}(C^q_{e''g}\rho_{eg} - C^q_{eg}\rho_{ge''})  \label{eq:rho_ee_lased}
\end{align}
\begin{align}
    \dot{\rho}_{ge} = & -i(\Delta^q_{eg}+\delta+\frac{1}{2\tau}+\frac{1}{2\tau_f}+\frac{1}{2\tau_b})\rho_{ge} \nonumber \\ 
    & - i\Omega\sum_{qe'}C^q_{e'g}\rho_{e'e} + i\Omega\sum_{qg'}C^q_{eg'}\rho_{gg'}  \label{eq:rho_ge_lased}.
\end{align}
The decay constants in $\dot{\rho}_{gg''}$ are contained in the term
\begin{align}
& \gamma^q_{e'e''}  =  \nonumber\\ & \begin{cases}
\frac{|C^q_{e'g''}C^q_{e''g}|}{\sum_{g'}|C^q_{e''g'}C^q_{e'g'}|}, & \text{if $e'= e''$}\\
\Gamma_{e'ge''g}, & \text{if $e'\not=e''$ AND $\sum_{q}C^q_{e'g}C^q_{e''g} \not= 0$}. \label{eq:gamma}
\end{cases}
\end{align}
where the sum over $q$ is the sum of all values over which spontaneous emission can occur: +1, 0, and -1. The second case in equation \ref{eq:gamma} only appears when there is hyperfine splitting leading to vertical coherences \cite{farrell1988quantum}. 

\begin{figure}[t]
    \centering
    \includegraphics[scale=0.6]{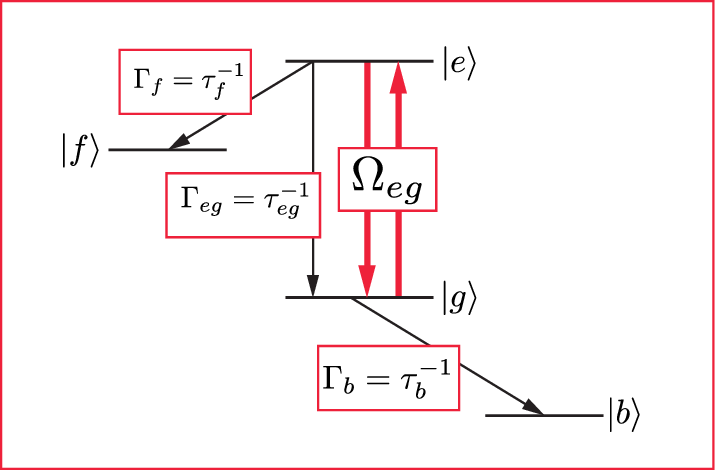}
    \caption{The states of the atomic system used in LASED can also include states $\ket{f}$ and $\ket{b}$ which are not directly coupled by the laser as shown. For details, see the text.}
    \label{fig:externaldecay}
\end{figure}

The calculation of the generalized decay constants and their phase is shown in appendix \ref{appendix:decay_const}. The coupling coefficients are calculated using equation \ref{eq:couplingcoeff} and the maximum half-Rabi frequency $\Omega$ is calculated using equation \ref{eq:rabifreq}. The detuning term is calculated using equation \ref{eq:doppler_detuning} as discussed in section \ref{section:eq_of_motion}. The laser-atom system modelled using equations \ref{eq:rho_gg_lased} to \ref{eq:rho_ge_lased} are more general than the system modelled using equations \ref{eq:rho_gg} to \ref{eq:rho_ge} as these also include extra decay terms that describe the process of relaxation to states which are not directly coupled by the laser, as shown in Fig. \ref{fig:externaldecay}. These include states that $\ket{e}$ and $\ket{g}$ may decay to that are not included in equations \ref{eq:rho_gg} to \ref{eq:rho_ge}, as well as any non-radiative decay routes that may occur. The decay from a laser-excited state $\ket{e}$ to a non-coupled state $\ket{f}$ is modelled by the lifetime $\tau_f$ and the decay from a lower state $\ket{g}$ to a non-coupled state $\ket{b}$ is modelled by the lifetime $\tau_b$. Equation \ref{eq:rho_ge_lased} also includes a detuning term $\delta$, which allows the user to add a constant detuning from resonance if required (e.g. for laser-cooling of atoms).  

Whilst the matrix $A$ is being generated the equations of motion can be printed out in a numeric or symbolic format depending on the user's preference. The Sympy package \cite{sympy2017} is used to generate the symbolic equations, which can be output as LaTeX.

Once the matrix $A$ has been generated, the NumPy package \cite{harris2020array} is used  to diagonalise the matrix to form $D$. NumPy is also used to perform all matrix multiplication in LASED. The SciPy package \cite{2020SciPy-NMeth} is used to generate the matrix of eigenvectors $V$ from $A$ and is also used in LASED to perform matrix exponentials and the inversion of matrices. For every element of the time array $t_i$ the column vector $\rho(t)$ of the laser-atom system is calculated numerically. Before looping over every element in $t_i$, $V^{-1}\rho(t_0)$ is calculated to save computation time. During the loop over $t_i$, the matrix exponential $e^{Dt}$ is calculated by taking the exponent of each diagonal element of $Dt$. Finally, $\rho(t)$ is calculated using equation \ref{eq:rho_matrix_exp}. Once the time evolution is completed, the user can access any element of $\rho(t)$ for analysis, the data can be saved as a csv file, or it can be plotted.

\subsection{Gaussian and Doppler Averaging} \label{section:gauss_and_dopp}
When a laser-atom system is modelled in LASED the default setting is that the spatial intensity profile of the laser beam is uniform. The user can however also specify a two-dimensional (2D) Gaussian laser beam profile, so as to emulate a TEM$_{00}$ mode \cite{welford1991efficient}. For a Gaussian beam the intensity as a function of the radial distance from its beam axis $r$ is given by
\begin{equation}
    I(r) = I_0e^{-\frac{r^2}{2r_{\sigma}^2}} \label{eq:gaussian_intensity}
\end{equation}
where $I_0$ is the intensity at the peak $r = 0$ and $r_{\sigma}$ is the radial distance equivalent to the 2D standard deviation. To obtain the total laser power $P_{\textrm{las}}$ as measured by a power meter, equation \ref{eq:gaussian_intensity} is integrated so that 
\begin{align}
    P_{\textrm{las}} = & \int^{\infty}_02\pi{r}I_0e^{-\frac{r^2}{2r_{\sigma}^2}}dr \nonumber \\
    & = 2\pi{r_{\sigma}^2}I_0
\end{align}
The intensity at any given radius is hence given by
\begin{equation}
    I(r) = \frac{P_{\textrm{las}}}{2\pi{r_{\sigma}^2}}e^{-\frac{r^2}{2r_{\sigma}^2}}. \label{eq:gaussian_intensity_full}
\end{equation}
from which the equivalent Rabi frequency can be generated. To model the effect of a Gaussian beam profile on the system, the beam profile is divided into a series of radial rings with the populations and coherences generated for each ring then summed incoherently to obtain the total density matrix for the ensemble. LASED assumes that the atoms are uniformly distributed throughout the laser beam profile with a density given by $\rho_{A}$ and that the atoms are stationary during the interaction. Hence the number of atoms in any ring between $r$ and $r + \Delta{r}$ is given by
\begin{equation}
    N^{\Delta{r}}_A = \rho_A(2\pi{r}\Delta{r}h) \label{eq:number_atoms_deltar}
\end{equation}
where the laser beam is assumed to be parallel through the interaction region, which has a height along the laser beam of $h$. Equation \ref{eq:number_atoms_deltar} then provides a weighting term to calculate the total number of atoms in the interaction region, up to a given radius. The laser beam diameter is approximated as 6$r_{\sigma}$ ($\pm3r_{\sigma}$) and so the total number of atoms in the interaction volume is given by 
\begin{equation}
    N^{6r_{\sigma}}_{\textrm{total}} = \int^{3r_{\sigma}}_0 2\pi\rho_Ahrdr = 9\pi\rho_Ahr_{\sigma}^2 \label{eq:number_of_atoms_6r}
\end{equation}
For $n_r$ equal rings and a beam diameter of 6$r_{\sigma}$ the ring radius will be $\Delta{r}=3r_{\sigma}/n_r$. The numerical calculation of the density matrix elements can then be calculated for averaging over the Gaussian laser profile, by performing a discrete sum of all $n_r$ rings and dividing by the total number of atoms:
\begin{align}
    \rho^{\textrm{av}}(I(r_{\sigma}), t) = & \frac{\sum^{n_r-1}_{j=0}\rho_A((2j+1)\pi\Delta{r^2}h)\rho(I(r_j, r_{\sigma}), t)}{9\pi\rho_Ahr_{\sigma}^2} \nonumber \\
    & =  \frac{\sum^{n_r-1}_{j=0}(2j+1)(\frac{9r_{\sigma}^2}{n_r^2})\rho(I(r_j, r_{\sigma}), t)}{9r_{\sigma}^2} \nonumber \\
    & = \frac{1}{n_r^2}\sum^{n_r-1}_{j=0} (2j+1) \rho(I(r_j, r_{\sigma}), t) \nonumber \\
    & =  \frac{1}{n_r^2}\sum^{n_r-1}_{j=0} (2j+1) \rho(\Omega(r_j, r_{\sigma}), t). \label{eq:gaussian_averaging_rho}
\end{align}
The half-Rabi frequency is introduced in place of the intensity in equation \ref{eq:gaussian_averaging_rho} since this is what is required when combining equations \ref{eq:rabifreq} and \ref{eq:gaussian_intensity_full}. To model a Gaussian beam profile in LASED, the user must enter the number of rings $n_r$ as well as the 2D standard deviation of the beam profile $r_{\sigma}$ in millimetres. When performing the time evolution, an array of ring radii is created up to the maximum beam profile radius 3$r_{\sigma}$. For each $r_j$ in the array, the time evolution of the laser-atom system is calculated and then averaged as given by equation \ref{eq:gaussian_averaging_rho}.

LASED also includes a functionality to model the effect of the Doppler profile of atoms within the interaction region, as would occur in an atomic beam from an oven or gas jet. The Doppler profile of the atoms is input to the model as a detuning term $\delta$ in units of 10$^9$ rad/s. For numerical purposes the Doppler profile is again split up into discrete values across the profile and the density matrix elements are calculated for each detuning term. The results are then averaged in a similar way to that adopted for representation of a TEM$_{00}$ laser beam. This Doppler averaging requires a weighting factor of the atoms given by \cite{siegman1986lasers}
\begin{equation}
    F_{\textrm{Dopp}}(\delta) = \frac{1}{\sqrt{2\pi\Delta_{Dopp}^2}}e^{-\frac{\delta^2}{2\Delta_{Dopp}^2}} \label{eq:doppler_gaussian_distr}
\end{equation}
where $\Delta_{\textrm{Dopp}}$ is the Doppler width. The averaged density matrix elements across the atomic Doppler profile are then given by
\begin{align}
    \rho^{\textrm{av}} & = \sum_i \rho({\delta_i})F_{\textrm{Dopp}}(\delta_i)\Delta\delta_i \nonumber \\
    & = \frac{1}{\sqrt{2\pi\Delta_{Dopp}^2}}\sum_i\rho(\delta_i)e^{-\frac{\delta^2}{2\Delta_{\textrm{Dopp}}^2}}\Delta\delta_i \label{eq:rho_average_doppler}
\end{align}
where $\Delta\delta_i$ is the angular frequency spacing between the discrete detunings that are used to represent the Doppler profile. Hence, to model a Doppler profile using LASED the user must declare a value for the Doppler width and create an array which contains discrete detuning values. Equation \ref{eq:rho_average_doppler} is then used to calculate the Doppler averaged density matrix elements for the system. 

\subsection{Rotation of quantization reference frames} \label{section:rotation}
It is often advantageous to define an atomic system in a particular reference frame that makes the calculation easier, or that decreases the computation time. As an example, excitation by linearly polarized light can adopt a quantization z-axis (QA) along the direction of the electric field vector, so that the change in  $m_{F}$ values between upper and lower substates is $\Delta{m_{F}}$ = 0. An alternative and equally valid representation for linear excitation may choose the quantization axis along the direction of the laser beam, in which case simultaneous $\Delta{m_{F}}$ = ±1 excitation occurs. In the former case for an $S$ to $P$ transition, this leads to n = 4 differential equations that must be solved. By contrast, in the latter case, nine equations must be generated and then solved. Both calculations lead to the same results and can be related to each other using a suitable rotation from one frame to the other. Since the computational speed scales as n$^2$, choosing the QA along the electric field vector in this example hence produces results more than 5 times faster than when the QA is chosen along the beam.

An example where the rotation technique has been adopted to simplify the calculation can be found in \cite{Stepwise_1990}, where electron excited mercury atoms in the 6$^1P_1$ state were further excited by a laser beam to the 6$^1D_2$ state using linearly polarized light. In this case the atomic system in the collision frame (QA along the direction of the electron beam) was first rotated into the laser frame along the electric field of the laser and the laser interaction was calculated in this new frame. The resulting atomic system was then rotated back to the collision frame to determine the evolved atomic structure in that frame. This required 36 differential equations to be solved for the laser interaction, compared to 64 equations that would need to be generated and solved simultaneously if the calculation had been carried out directly in the collision frame. A further advantage of moving to the laser frame was that the 12 equations for the populations and optical coherences decoupled from the 24 equations for the non-optical and atomic coherences, so that the matrix $A$ was block diagonal. This lead to a 5.7 fold increase in computational efficiency.

It is not however always possible to apply this technique, since there may be constraints on the system due to additional interactions. An example is found in \cite{ murray2008theoretical,superelasticBfieldPRL,superelasticBfieldJPB}, where an external magnetic \textbf{B}-field was imposed on the system. In these experiments the  \textbf{B}-field direction was co-linear with the direction of the laser beam, and so the QA was chosen along this axis for excitation by both circular and linearly polarized laser beams, with the linear beam being considered as a superposition of right-hand and left-hand circularly polarized beams.  

LASED can incorporate rotation between reference frames within its structure, so that these advantages can be exploited. The rotation is performed by rotating the density matrix for each atomic state using the Wigner rotation matrices \cite{brink1963angular}, so that
\begin{equation}
    \rho_{Jm,J'm'} = \sum_{\mu = -J}^{+J} \sum_{\mu'=-J'}^{+J'} D^{J*}_{\mu{m}}(\omega)\rho_{J\mu,J'\mu'}D^{J'}_{\mu'm'}(\omega) \label{eq:wigner_D_rotation_rho}
\end{equation}
where $\rho_{J\mu,J'\mu'}$ and $\rho_{Jm,J'm'}$ are the atomic state density matrix elements in the new and old reference frame respectively, $J$ is the total angular momentum of the state (which will be $F$ if there is non-zero isospin), $m$ is the projection of angular momentum onto the QA, and $\omega$ denotes the Euler angles for the rotation $(\alpha, \beta, \gamma)$. In LASED, the Euler angles are defined as three angles of rotation performed in succession from Cartesian reference frame $Z$ to $Z'$, and then to a final $Z''$. $\alpha$ then rotates around the z-axis, $\beta$ rotates around the new y$'$-axis, and $\gamma$ finally rotates around the new z$''$-axis. The Wigner-D matrix is calculated using \cite{biedenharn1981angular}
\begin{equation}
    D^J_{m'm}(\omega) = e^{-im'\alpha}d^J_{m'm}(\beta)e^{-im\gamma} \label{eq:wigner_D}
\end{equation}
where $d$ is determined using
\begin{align}
    & d^J_{m'm}(\beta) =  \sqrt{(J+m')!(J-m'!)(J+m)!(J-m)!} \nonumber \\
    & \times \sum^{s_{\textrm{max}}}_{ s=s_{\textrm{min}}}\frac{(-1)^{m'-m+s}(\frac{\textrm{cos}\beta}{2})^{2J+m-m'-2s}(\frac{\textrm{sin}\beta}{2})^{m'-m+2s})}{(J+m-s)!s!(m'-m+s)(J-m'-s)!} 
\end{align}
The summation over $s$ is constrained to $s_{\textrm{min}} = \textrm{max}(0, m-m')$ and $s_{\textrm{max}} = \textrm{min}(J+m, J-m')$ so that the factorials remain non-negative. Hence, if a rotation matrix is required to rotate a state with angular momentum $J$, it will be a square matrix of size $2J+1$. If required, LASED uses equation \ref{eq:wigner_D_rotation_rho} to rotate any density matrix set up by the user to a new reference frame.

\subsection{Modelling different laser polarizations}
\label{section:polarization}

    \begin{figure}[b]
    \centering
    \includegraphics[scale=1.0]{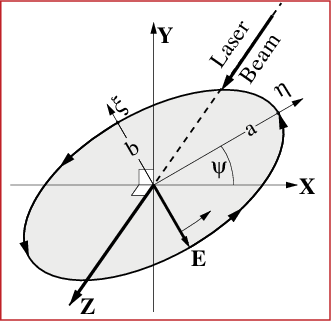}
    \caption{Co-ordinate system for elliptically polarized light propagating in the +z-direction. The electric field vector \textbf{E} traces out an ellipse with a major axis $a$ and minor axis $b$. The major axis is at an angle $\psi$ with respect to the co-ordinate system shown. The vector rotates in an anti-clockwise direction in this example and so is right-hand polarized.}
    \label{fig:ellipticlight}
\end{figure}

In many experiments the laser beam interacting with the atoms is chosen to have either circular or linearly polarization. This makes the generation of the equations of motion and subsequent computation of the dynamics relatively straightforward, as discussed above. It is also important for LASED to model the interaction using a laser which has elliptic polarization, since this is the most general form for any beam. An elliptically polarized beam can be considered as one that has its \textbf{E}-field vector tracing out an ellipse, as shown in Fig. \ref{fig:ellipticlight}. The ellipse has major and minor axes, with the major axis being rotated from the x-axis at an angle $\psi$ as shown. The direction of rotation of the \textbf{E}-field also must be defined to fully characterize the radiation.

Any elliptically polarized beam can be described as a superposition of right-hand and left-hand circular components with different complex amplitudes, the relative phase between the amplitudes producing the rotation of the major axis from the x-axis. These amplitudes then feed into the Rabi frequencies through equations \ref{eq:couplingcoeff} and \ref{eq:rabifreq}. Since the QED model has been developed for the Rabi frequencies being real, it is necessary to first rotate the QA through the angle $\psi$ so that the new x-axis is aligned along the major axis of the ellipse. This rotation sets the relative phase to zero and so the elliptically polarized light can then described using two real amplitudes, as given by equation \ref{eq:elliptical_polarization_ket}. 

\begin{figure*}[ht]
    \centering
    \includegraphics[scale=0.57]{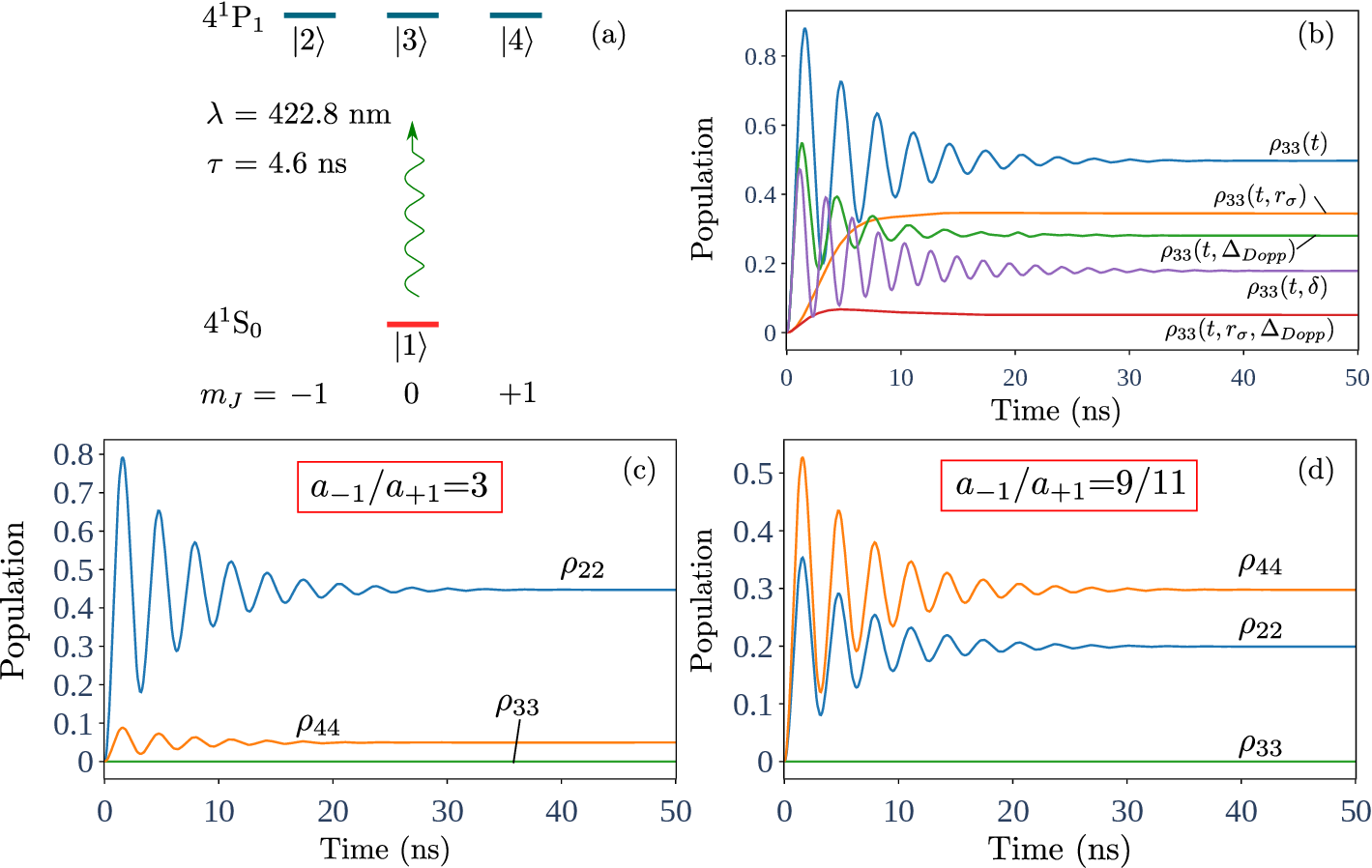}
    \caption{(a) A level diagram for the 4$^1S_0$ to 4$^1P_1$ excitation of calcium using linearly polarized light, with the QA co-linear with the \textbf{E}-field of the laser beam so that $\Delta{m_J} = 0$. (b) The simulated time evolution of the population of substate $\ket{3}$, showing the effect of various simulation parameters, including a Doppler profile for the atomic beam and a Gaussian laser profile (see section \ref{subsection:CalciumSP} for details). (c) and (d) show the excited state populations using elliptically polarised light for different weightings $a_{-1}$ and $a_{+1}$. In this case the atom is described with the QA along the laser beam direction and so substate $\ket{3}$ remains unpopulated. All simulations used a laser intensity of 100 mW/mm$^2$.}
    \label{fig:ca}
\end{figure*}

\begin{equation} \label{eq:elliptical_polarization_ket}
    \ket{P} = \frac{1}{\sqrt{a_{-1}^2+a_{+1}^2}}(a_{-1}\ket{\sigma_{-1}}+a_{+1}\ket{\sigma_{+1}})
\end{equation}
Here $\ket{\sigma_{-1}}$ and $\ket{\sigma_{+1}}$ are the LHC and RHC polarization unit vectors and $a_{-1}$ and $a_{+1}$ are real amplitudes. The computation then proceeds in the same way as described above, however two weighted Rabi frequencies are now required to describe the interaction. From equation \ref{eq:elliptical_polarization_ket} it follows that the half-Rabi frequency for elliptically polarised light in this frame is given by
\begin{equation} \label{eq:elliptical_rabi_freq}
    \Omega_{\textrm{elliptic}} = \frac{1}{\sqrt{a_{-1}^2+a_{+1}^2}}(a_{-1}\Omega_{-1}-a_{+1}\Omega_{+1})
\end{equation}
where the negative sign arises from the definition of the dipole moment in a circular basis. Once the interaction has been modelled in this frame, the QA can be rotated back into the original frame to calculate the final density matrix elements.

LASED uses the procedure detailed above to model excitation by elliptically polarized light, if this is required. The user can enter any polarization state into the model, however they must also include the normalisation factor to ensure the correct Rabi frequency is calculated. As an example, if the minor axis of the ellipse has $b = 0$, the ellipse represents linearly polarized light and so the half-Rabi frequency in this frame is represented by an equal weighting of the circular basis states. In this case $a_{-1} = a_{+1} = 1$ and so the normalisation factor to be input is $1/\sqrt{2}$.

\begin{figure*}[ht]
    \centering
    \includegraphics[scale=0.6]{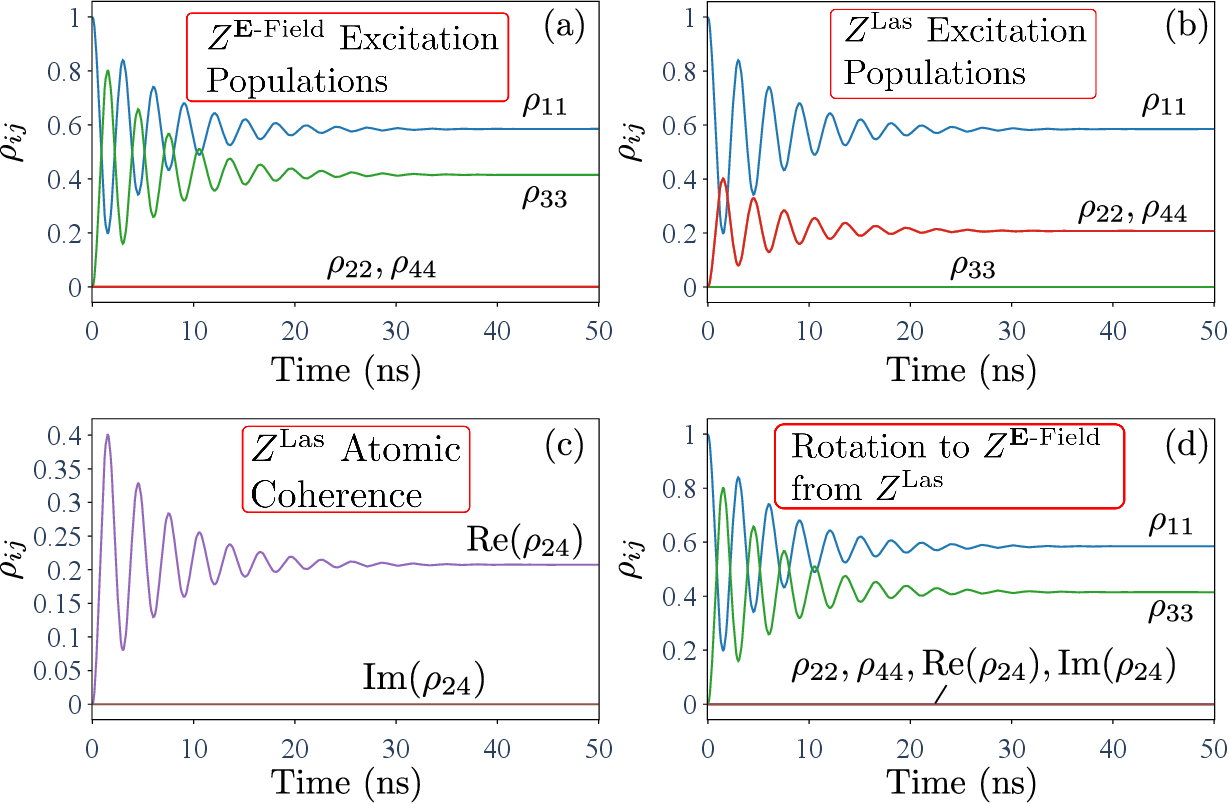}
    \caption{Figure showing the use of rotations in LASED to check that the general equations of motion are consistent in all reference frames using the calcium system described in Fig. \ref{fig:ca}(a). In panel (a) excitation occurs in the frame $Z^{\textrm{\textbf{E}-Field}}$ with the QA along the \textbf{E}-field of the linearly polarized laser beam. An intensity of 100 mW/mm$^2$ is used and the laser is detuned by 100 MHz from resonance. The populations of the states $\ket{1}$ and $\ket{3}$ in the frame are shown. In (b) excitation is now using simultaneous $\sigma^{-}$ and $\sigma^{+}$ radiation with the reference frame $Z^{\textrm{Las}}$ QA along the laser beam direction. The populations $\rho_{11}$, $\rho_{22}$ and $\rho_{44}$ are shown. Panel (c) shows the real and imaginary components of the atomic coherence $\rho_{24}$. The data from (b) and (c) are then rotated back to $Z^{\textrm{\textbf{E}-Field}}$ in panel (d). This produces results identical to those in panel (a).}
    \label{fig:Ca_rot}
\end{figure*}

\subsection{Visualising the shape of the charge cloud}
\label{section:chargecloud}

In LASED, the three dimensional angular shape of the charge cloud for the lower and upper states can be visualised as given in \cite{masters1996multipole,murray2008theoretical}, using the expression
\begin{equation}
    W(\theta, \phi,t) = \sum_{mm'} \rho_{Jm,Jm'}(t) Y_{Jm}(\theta, \phi)Y^*_{Jm'}(\theta, \phi) \label{eq:angular_shape}
\end{equation}
 where $Y_{Jm}$ are spherical harmonics, $J$ is the total angular momentum of the state and $m$ is the projection of $J$ onto the selected quantization axis. $\rho_{mm'}(t)$ is the time dependent density matrix element for the atomic state that is being visualised. In LASED, the user can generate the angular shape of the states W($\theta$,$\phi$,$t$) in the laser-atom system as the system evolves over time. Images of the charge cloud can then be created using any plotting package. These images can then be displayed sequentially as a function of time, using software that creates a video from the image sequence. The generated videos can be instructive to demonstrate how the states evolve under different experimental conditions. Examples of the generated charge clouds at different times for both the lower and upper states are shown in Fig. \ref{fig:3D_10P_populations_he} in section \ref{section:examples}.

\section{Examples from LASED modelling} \label{section:examples}

In this section the features of LASED are presented using examples from different laser-atom systems. For each simulation, the time evolution of the populations of various atomic substates is presented. The sum of the populations of all substates is initially set to unity and so the populations directly represent the probability of a particular atom in the ensemble being in that substate at any given time. Additional examples using LASED can be found at \cite{LASEDreadthedocs}.

\subsection{Calcium $\textbf{S}$ to $\textbf{P}$ Excitation}  \label{subsection:CalciumSP}

The simplest system to simulate is from an $S$ state to a $P$ state and so as an example, laser excitation from the 4$^1S_0$ to the 4$^1P_1$ state in calcium is considered. A level diagram is shown in Fig. \ref{fig:ca}(a) for this transition. The lifetime and transition wavelength are taken from \cite{lurio1964lifetime} and \cite{risberg1968spectrum} respectively. In Fig. \ref{fig:ca}(b) the time evolution of the upper state population $\rho_{33}$ is presented under different conditions. These include a fixed laser detuning of $\delta = 300 \textrm{ MHz}$, a Doppler atomic beam profile with $\Delta_{\textrm{Dopp}} = 300 \textrm{ MHz}$, a Gaussian laser beam profile with $P_{\textrm{las}} = 100 \textrm{ mW}$ and $r_{\sigma}$ = 0.75 mm, and when both Doppler and Gaussian averaging processes are included together. The simulation time was from 0 to 50 ns using 501 time steps. These simulations are in agreement with the calculations presented in \cite{murray2003low}. 

Results for the same system with elliptically polarized light are shown in Fig. \ref{fig:ca}(c) and (d). In panel (c) the weightings are set to $a_{-1}$/$a_{+1}$ = 3.0 whereas in panel (d) $a_{-1}$/$a_{+1}$ = 0.8. As expected, the population of the $m_J$ = -1 state is much larger in panel (c) due to the favoured $\ket{\sigma_{-1}}$ weighting. By contrast in panel d) where the weighting for the $\ket{\sigma_{+1}}$ basis state is higher, the $m_J = +1$ substate population dominates. The population of substate $m_J$ = 0 is identically zero for the entire simulation as the laser cannot couple to state $\ket{3}$ in this frame with $\sigma^+$ and $\sigma^-$ polarization. 

An example of using rotations in LASED can be seen in Fig. \ref{fig:Ca_rot}. To check that LASED is valid in all reference frames the calcium system described in Fig. \ref{fig:ca}(a) is once again considered. This system is now excited using linear-polarised light with the QA along the \textbf{E}-field of the laser, for a laser intensity of 100 mW/mm$^2$ and a detuning of 100 MHz. The results from this simulation are shown in panel \ref{fig:Ca_rot}(a). Under these conditions four differential equations are required to describe the populations of substates $\ket{1}$ and $\ket{3}$ as well as the optical coherences generated between them. An equally valid representation is to choose the QA along the direction of the laser beam. In this frame substates $\ket{2}$ and $\ket{4}$ are excited using simultaneous $\sigma^{-}$ and $\sigma^{+}$ radiation. Substate $\ket{3}$ in this frame remains unpopulated. In this representation nine equations must be generated and solved. Three equations represent the populations of substates $\ket{1}$, $\ket{2}$ and $\ket{4}$, four equations represent the optical coherences between them and two equations represent the atomic coherences generated between substates $\ket{2}$ and $\ket{4}$.

\begin{figure*}[ht]
    \centering
    \includegraphics[scale=0.65]{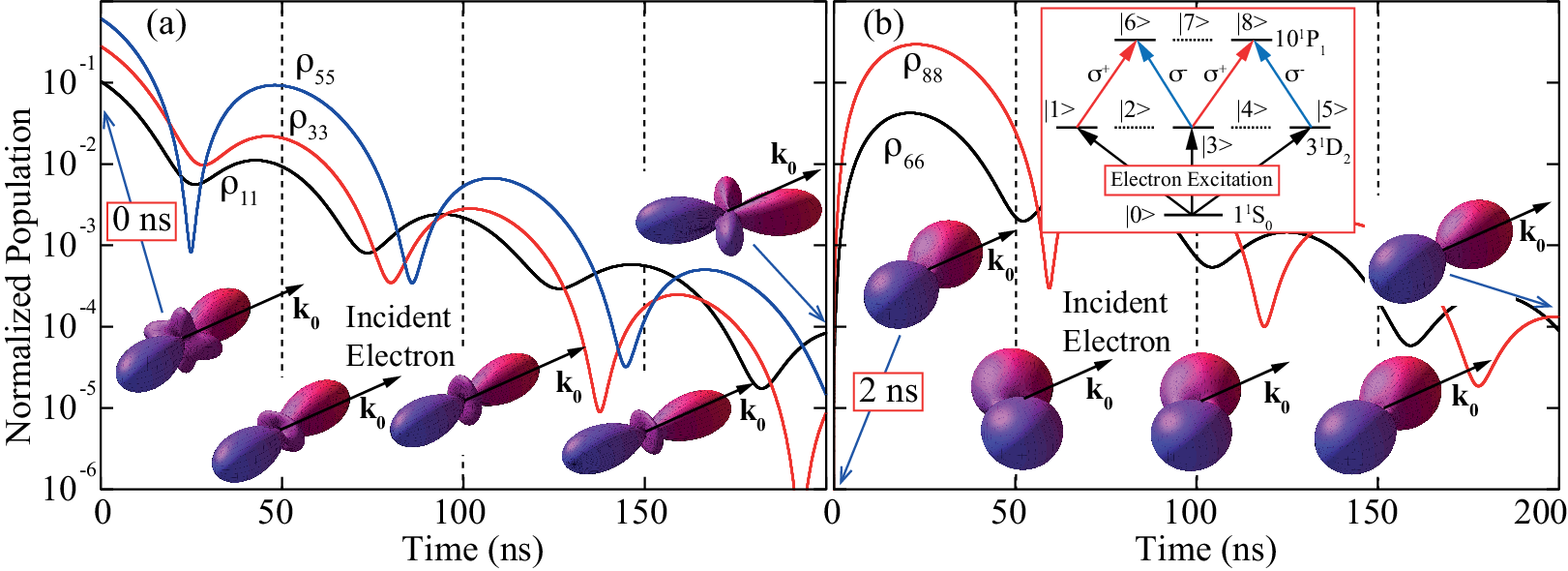}
    \caption{Example of stepwise electron and laser excitation from the 3$^1D_{2}$ state of helium to the 10$^1P_{1}$ state, represented in the Natural Frame where the QA is orthogonal to the scattering plane. The 3$^1D_{2}$ state is excited by electron impact and is stepwise-excited to the 10$^1P_{1}$ state, as shown in the inset figure. Both 3$^1D_{2}$ and 10$^1P_{1}$ states can decay to states that are not coupled by the laser as discussed in the text. (a) shows the time evolution of the 3$^1D_{2}$ state populations and (b) shows that of the 10$^1P_{1}$ state, both presented on a logarithmic scale. The structure of the associated charge clouds have been calculated at various times during the evolution of the states as shown. This is given for the 10$^1P_{1}$ state at 2 ns, since this state is unpopulated at 0 ns when the laser is switched on. The incident electron direction \textbf{k}$_{0}$ is also shown for reference. The atomic coherences are not shown, however they are calculated in LASED to allow the charge cloud models to be generated. }
    \label{fig:3D_10P_populations_he}
\end{figure*}

The results from this calculation are shown in Fig. \ref{fig:Ca_rot}(b) for the populations $\rho_{11}$, $\rho_{22}$ and in Fig. \ref{fig:Ca_rot}(c) for the atomic coherence $\rho_{24}$. Note that Im($\rho_{24}$) = 0 here due to the choice of axes in both reference frames. The results from this calculation are then rotated back to the reference frame where the QA is along the \textbf{E}-field of the laser in Fig. \ref{fig:Ca_rot}(d), which reproduces the results in Fig. \ref{fig:Ca_rot}(a) exactly. This shows that LASED produces the same result independent of the reference frame chosen, as long as the initial conditions are rotated before excitation.  

\begin{figure}[hb]
    \centering
    \includegraphics[scale=0.35]{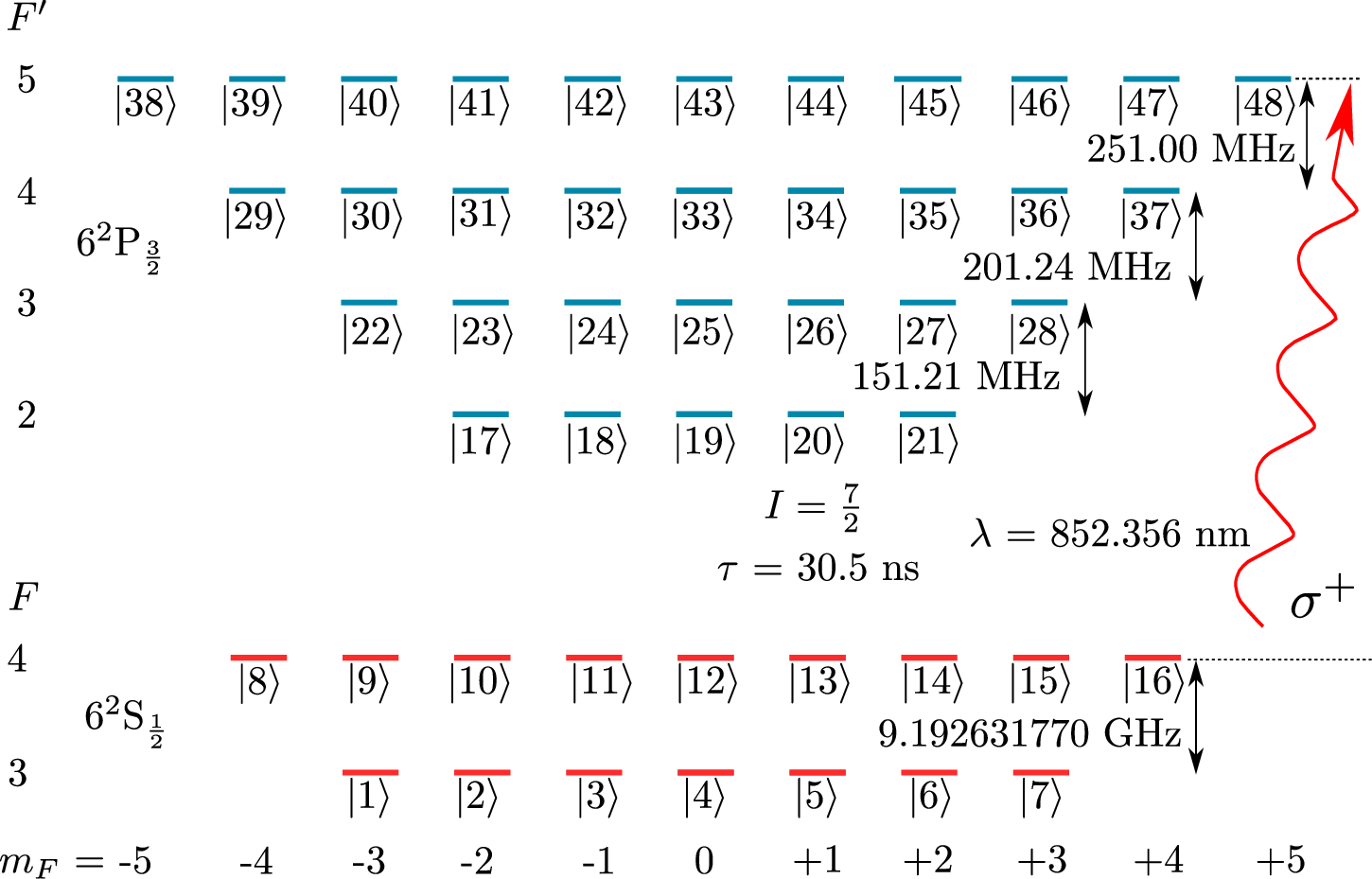}
    \caption{A level diagram of the 6$^2$S$_{1/2}$ to 6$^2$P$_{3/2}$ excitation in caesium. $\sigma^+$ exciting laser radiation is represented by a wiggly arrow and is set to be on-resonance between the F = 4 lower state and the $F'$ = 5 upper state.}
    \label{fig:caesium_level_scheme}
\end{figure}

\begin{figure*}[ht]
    \centering
    \includegraphics[scale=0.75]{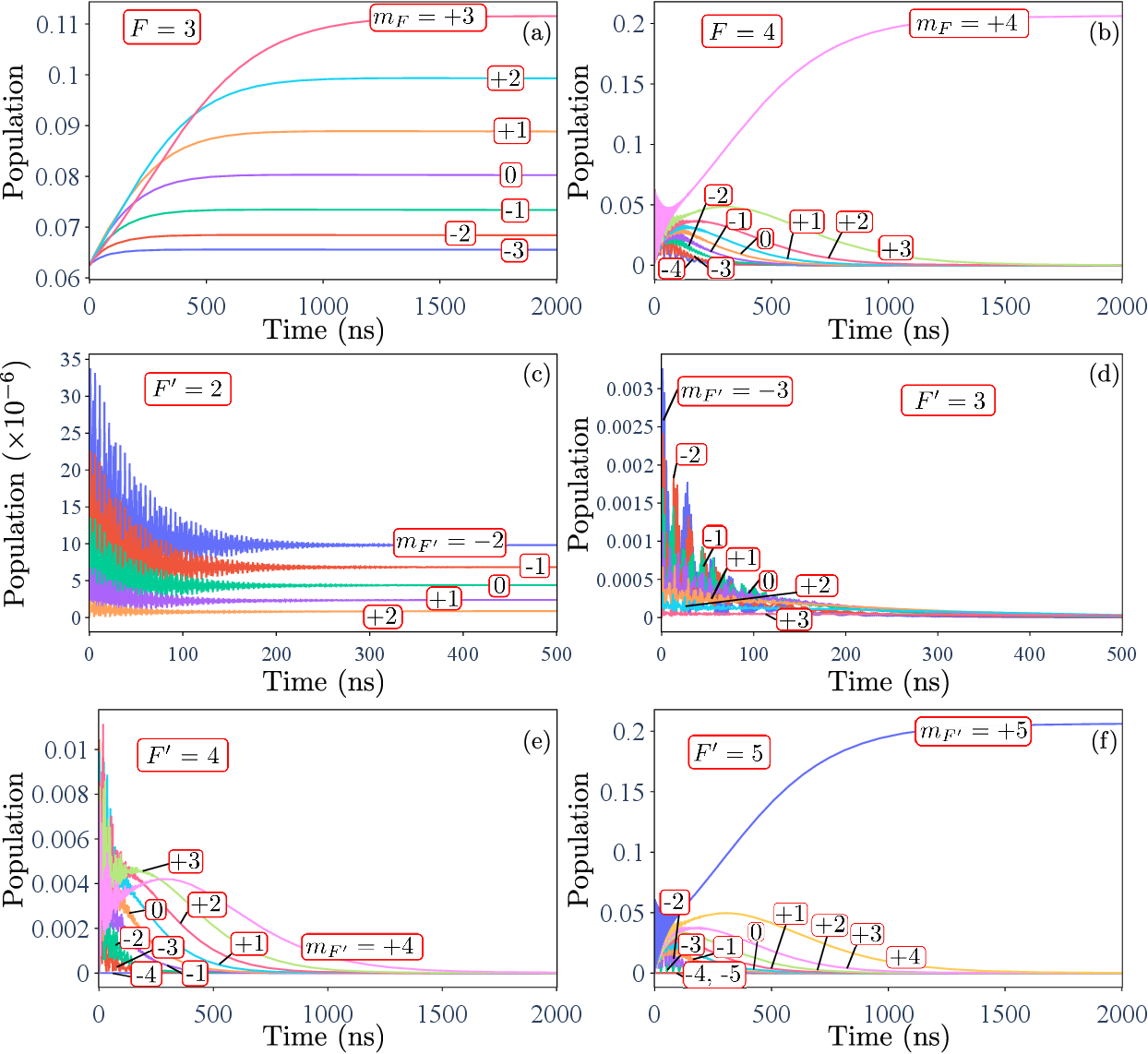}
    \caption{\label{fig:caesium_evolution}The time evolution of the populations of substates in the ground states (a) F = 3, (b) $F$ = 4, the upper states (c) $F'$ = 2, (d) $F'$ = 3, (e) $F'$ = 4, and (f) $F'$ = 5 for the system described in Fig. \ref{fig:caesium_level_scheme}. The laser is $\sigma^+$-polarised with an intensity of 50 mW/mm$^2$.}
\end{figure*}

\subsection{3$^1\textbf{D}_{2}$ to 10$^1\textbf{P}_{1}$ excitation in helium following electron impact from the 1$^1\textbf{S}_{0}$ state }
\label{section:helium}

A more complex system to model using LASED is presented in this section, where laser excitation is from the 3$^1D_{2}$ state to the 10$^1P_{1}$ state, as shown in the inset in Fig. \ref{fig:3D_10P_populations_he}. This transition is of interest as experiments are in preparation in Manchester to study this stepwise excitation process. Excitation from the the 1$^1$S$_{0}$ state to the 3$^1D_{2}$ state is via electron collision and so the system is presented in the Natural frame $Z^{\textrm{Nat}}$, where the QA is set orthogonal to the scattering plane spanned by the ingoing electron momentum \textbf{k}$_{0}$ and the outgoing electron momentum \textbf{k}$_{1}$. The laser beam is then injected along the quantization axis and is linearly polarized along the incident beam direction, with an incident intensity of 1,500 mW/mm$^2$. The beam is set to be on-resonance with the transition at a vacuum wavelength of 899.75205 nm. 

In this frame, only the substates $m_{J}$ = -2, 0, +2 in the 3$^1D_{2}$-state are excited due to reflection symmetry in the scattering plane \cite{beijers1987polarisation}. Since the laser beam is linearly polarised the interaction must be represented by simultaneous $\sigma^+$ and $\sigma^-$ excitation with equal weighting in this frame, as discussed above.

The initial 3$^1D_{2}$ substate populations and atomic coherences at t = 0 ns are taken from the Convergent Close Coupling calculation of Bray at Curtin University \cite{IgorBrayprivate}, for an electron impact energy of 40 eV and a scattering angle of 45$\degree$. The collision excites the atom into a superposition of $\ket{J,m}$ eigenstates $\ket{2,-2} = \ket{1}$, $\ket{2,0} = \ket{3}$, and $\ket{2,+2} = \ket{5}$. The states $\ket{2,-1} = \ket{2}$ and $\ket{2,+1} = \ket{4}$ are not initially populated. Simultaneous $\sigma^{+}$ and $\sigma^{-}$ laser excitation is used to represent linear excitation in the Natural frame, the 10$^1P_{1}$ states being excited by $\Delta{m_J}$ = +1 and -1 radiation so as populate the eigenstates $\ket{1,-1}$ = $\ket{6}$ and $\ket{1,+1}$ = $\ket{8}$, with the state $\ket{1,0}$ = $\ket{7}$ remaining unpopulated. Both 3$^1D_{2}$ and 10$^1P_{1}$ states can decay to states that are not coupled by the laser. The excited 3$^1D_{2}$ state decays to lower states $\ket{b}$ with a total lifetime of 15.7 ns. The upper 10$^1P_{1}$ state decays to states $\ket{f}$ with a lifetime of 59.6 ns, whereas the lifetime for decay back to the 3$^1D_{2}$ state is 80.7 $\mu$s \cite{NIST_ASD}. These decay routes are not shown in the inset of Fig. \ref{fig:3D_10P_populations_he} for clarity. 

Since the decay routes to $\ket{b}$ and $\ket{f}$ are relatively rapid, the populations of the 3$^1D_{2}$ and 10$^1P_{1}$ states are presented on a logarithmic scale in Fig. \ref{fig:3D_10P_populations_he}. The atomic and optical coherences are not shown, however these are also calculated by LASED. The decay routes to $\ket{f}$ and $\ket{b}$ leak both populations and atomic coherences away from the system. The population $\rho_{77}$ for the upper 10$^1P_{1}$ state remains zero throughout the simulation, since the laser does not couple to this state in the Natural frame. 

The full density matrices representing the 3$^1D_{2}$ and 10$^1P_{1}$ states are calculated by LASED, including the time evolution of both populations and atomic coherences. This allows the charge clouds associated with each state to be modelled as a function of time, as discussed in section \ref{section:chargecloud}. Examples of these charge cloud models are shown at different times throughout the evolution of the system in Fig. \ref{fig:3D_10P_populations_he}, for both the 3$^1D_{2}$ state and the 10$^1P_{1}$ state. It is seen that both the angle and shape of the $D$-state and P-state charge clouds evolve in a complex way, and so must be considered carefully in the associated experiments studying this system.

\subsection{Caesium $\textbf{D}_2$-Line}

LASED can also simulate the time evolution of systems with hyperfine structure, such as the caesium transition from the 6$^2S_{1/2}$ state to the 6$^2P_{3/2}$ state, commonly called the $D_2$-line. A level diagram for this system is shown in Fig. \ref{fig:caesium_level_scheme}. In this example, the laser is on resonance between the $F$ = 4 and $F'$ = 5 states and is set to have $\sigma^+$ polarization. The lifetime and wavelength of this transition are taken from \cite{rafac1999fast} and \cite{udem2000absolute} respectively. The hyperfine splittings for the upper and lower substates are taken from \cite{arimondo1977experimental}. 

The time evolution of a subset of the populations in the caesium manifold described in Fig. \ref{fig:caesium_level_scheme} is shown in Fig. \ref{fig:caesium_evolution}. This simulation was run with all ground states populated equally at $t$ = 0 ns and with no atomic coherences in the initial state, as would be produced for atoms emitted from an oven. Since the laser is tuned from the $F = 4$ state, selection rules prohibit excitation from this state to the $F'$ = 2 state, which hence remains essentially unpopulated as shown in Fig. \ref{fig:caesium_evolution}(c). The very small change in the population of these states as seen in Fig. \ref{fig:caesium_evolution}(c) arises due to pumping from the $F$ = 3 state by the laser radiation red-detuned by 9,193 MHz. Selection rules allow the states $F'$ = 3 and $F'$ = 4 to be populated from the $F$ = 4 state, however since they are detuned from resonance by 251 MHz and 452.24 MHz respectively, they are excited with only a small probability as the interaction proceeds. 

In Fig. \ref{fig:caesium_evolution}(a) the population evolution over time of the $F$ = 3 ground state is shown. This is the lowest state in the system and is not coupled directly by the laser beam. Since the $F'$ = 4 and $F'$ = 3 states can however decay to this state via spontaneous emission, its population slowly increases with time as shown, until the substates reach a steady state at around 1500 ns. As the interaction progresses, the states to the left of figure  \ref{fig:caesium_level_scheme} decrease in population since spontaneous emission feeds their population to the right, due to pumping with $\sigma^+$ radiation. This feeding to the right and subsequent decrease in population is seen in the substates of the $F'$ = 3 and $F'$ = 4 upper states in Figs. \ref{fig:caesium_evolution}(d) and \ref{fig:caesium_evolution}(e). Each of these states is effectively emptied within around 1500 ns. By contrast, the populations of the $F'$ = 5, $m_{F'}$ = +5 and F = 4, $m_{F}$ = +4 substates shown in Figs. \ref{fig:caesium_evolution}(b) and \ref{fig:caesium_evolution}(f) are seen to rise steadily after the Rabi oscillations have decayed, which occurs at around 150 ns. The substate populations rise quickly until they reach a steady state, after which they remain unchanged. This is a direct consequence of the system evolving towards the closed 2-level system between substates $\ket{16}$ and $\ket{48}$. After this time the interaction can then be approximated to a 2-level system between these substates, with spontaneous and stimulated emission from $\ket{48}$ always feeding back into substate $\ket{16}$. This simplified system is often used to simulate laser interactions in atom cooling and trapping experiments in a Magneto Optical Trap (MOT). 

LASED can easily simulate these large and complex systems and can generate all the equations of motion that are required. As noted above the computation time increases considerably as the number of states increases. As an example, simulation of the the calcium system in Fig. \ref{fig:ca} required less than 1 second of computing time. The model for the $D$ to $P$ state transition in figure \ref{fig:3D_10P_populations_he} took a few seconds to generate the data. By contrast, the caesium system required around 9,000 seconds of computing time to generate the results shown in Fig. \ref{fig:caesium_evolution}. 

\section{Conclusion} \label{section:conclusion}

LASED is an open source package available to researchers, that is written in the python programming language. The general equations of motion used in LASED have been described in this paper. LASED allows the user to model different aspects of the interaction, including the Doppler profile of an atomic beam, the Gaussian profile of a TEM$_{00}$ laser beam, an arbitrary polarization of the beam, any rotation between different frames of reference, and the angular shape of the atomic electron cloud. Examples of these techniques have been described here, using different atomic systems.

The purpose of LASED is to be a general, easy-to-use laser-atom system simulator which can be used for any atomic system excited by laser light. Later versions of the LASED library aim to extend its usefulness by including modelling of the interaction in magnetic fields, as well as including excitation by multiple laser beams. Computation times can be reduced by carefully considering the symmetry of the system and by eliminating equations that represent density matrix elements that remain zero throughout the simulation. In future versions of LASED, computationally intensive tasks such as generating the matrix $A$ will be implemented in the C++ programming language whilst still maintaining the python programming-interface. This will greatly increase speed. To further extend the usefulness of LASED the authors have made this package freely accessible, so that other researchers can contribute to its development and further extend its functionality. The source code is hence available and can be edited at \cite{LASEDgithub}. 

\section{Acknowledgements}
We wish to thank the Engineering and Physical Sciences Research Council (EPSRC) for funding through grants R120272, R125924 and R126554. Manish Patel would like to thank the University of Manchester for providing a PhD scholarship to carry out this work. We would also like to thank Professor Igor Bray for providing unpublished data for the electron impact excitation of helium, as used in section \ref{section:helium}.

\appendix

\section{Computing time}
LASED aims to model an arbitrary atom-laser system as defined by the user. By designing LASED to be as general as possible, the computing time increases rapidly with the number of substates in the system. At the same time, LASED aims to be efficient and usable with low-powered machines on any operating system. The python language was hence chosen for its development as this is open source and can be run on a wide range of different platforms. 

\begin{table}[hb]
\begin{tabular}{cc}
\hline
\multicolumn{1}{l}{n (Number of Energy Levels)} & \multicolumn{1}{l}{Execution Time (s)} \\ \hline
4 & 0.833 \\
6 & 0.855 \\
8 & 2.50 \\
24 & 283 \\
36 & 2270 \\
48 & 8570 \\ \hline
\end{tabular}
\caption{A table to show execution times using LASED to simulate the time evolution of laser-atom systems with varying number of energy levels. 501 time steps were simulated from 0 to 500 ns with a laser intensity of 100 mW/mm$^2$, $\pi$-polarised light, and no Gaussian or Doppler averaging. For n$\geq$24 the simulated systems have hyperfine structure.}
\label{table:execution_times}
\end{table}

To illustrate the computation time on a standard PC, table \ref{table:execution_times} shows the execution time for atomic systems that have different energy levels and substates. These computation times were obtained using an Intel i5-3320M CPU operating at 2.60 GHz using a Linux operating system with 8 GB of RAM.

\section{Calculation of the generalized decay constants} 
\label{appendix:decay_const} 

The generalized decay constants need to be calculated directly when there are vertical coherences in a laser-atom system i.e. when there is hyperfine splitting, as shown in equation \ref{eq:gamma}. For hyperfine states, the splitting between excited energy levels is small so $\omega_{e'} \approx \omega_{e''}$. Using equation \ref{eq:Gamma_egeg} this approximation hence leads to

\begin{equation}
    \Gamma_{ege'g} =  2\sum_{q} g^q_{e'g}g^{q*}_{eg}\pi\delta(\omega_q-\Delta_{eg})
\end{equation}
and from equation \ref{eq:Gamma_eg} the magnitude of the generalized decay constant can be calculated by
\begin{equation}
    |\Gamma_{ege'g}| = \sqrt{\Gamma_{eg}\Gamma_{e'g}}.
\end{equation}
The sign of the generalized decay constant is calculated by considering the coupling coefficients. The coupling coefficients are generally complex
\begin{align}
    & g^q_{e'g} = |g^q_{e'g}|e^{i\alpha} \\
    & g^q_{eg} = |g^q_{eg}|e^{i\zeta}
\end{align}
and the half-Rabi frequency can be written in terms of phase and amplitude terms, so that 
\begin{align}
    \Omega^q_{e'g} = g^q_{e'g}\langle{a_L(0)}\rangle & = |g^q_{e'g}|e^{i\alpha}\langle{a_L(0)}\rangle{e^{i\beta}} \nonumber \\
    & = |g^q_{e'g}|\langle{a_L(0)}\rangle{e^{i(\alpha+\beta)}}
\end{align}
and similarly 
\begin{equation}
    \Omega^q_{eg} =  |g^q_{eg}|\langle{a_L(0)}\rangle{e^{i(\zeta+\beta)}}.
\end{equation}
Since the half-Rabi frequencies are defined here as being real, it follows that 
\begin{align}
    \alpha + \beta = n\pi \label{eq:alpha+beta}\\
    \zeta + \beta = m\pi \label{eq:zeta+beta}
\end{align}
where $n$ and $m$ are integers. If equation \ref{eq:alpha+beta} is subtracted from \ref{eq:zeta+beta} it is found that
\begin{equation}
    e^{i(\alpha-\zeta)} = e^{i(n-m)\pi} = \begin{cases}
    +1, \text{if $n-m$ even} \\
    -1, \text{if $n-m$ odd}.
\end{cases}
\end{equation}
Hence if $\Omega^q_{eg}$ and $\Omega^q_{e'g}$ have the same sign then $(n-m)$ is even and if they have the opposite sign then $(n-m)$ is odd. This can be related to the calculated coupling coefficients using equation \ref{eq:rabifreq} so that 
\begin{equation}
    \Gamma_{ege'g} = \begin{cases}
    +|\Gamma_{ege'g}|, \text{if $C^q_{eg}C^q_{e'g'} > 0$} \\
    -|\Gamma_{ege'g}|, \text{if $C^q_{eg}C^q_{e'g'} < 0$}.
    \end{cases}
\end{equation}

\section{Installation of LASED}
\label{appendix:installation}
Installation of LASED requires the user to install the python programming language which can be found at \cite{python_download}. It is recommended to download and install the latest source release of python. Once python has been installed, the command "pip3 install LASED" must be input to the terminal and run. This will install LASED and all dependencies. 

\bibliography{bibliography.bib}

\providecommand{\noopsort}[1]{}\providecommand{\singleletter}[1]{#1}%
\begin{thebibliography}{51}%
\makeatletter
\providecommand \@ifxundefined [1]{%
 \@ifx{#1\undefined}
}%
\providecommand \@ifnum [1]{%
 \ifnum #1\expandafter \@firstoftwo
 \else \expandafter \@secondoftwo
 \fi
}%
\providecommand \@ifx [1]{%
 \ifx #1\expandafter \@firstoftwo
 \else \expandafter \@secondoftwo
 \fi
}%
\providecommand \natexlab [1]{#1}%
\providecommand \enquote  [1]{``#1''}%
\providecommand \bibnamefont  [1]{#1}%
\providecommand \bibfnamefont [1]{#1}%
\providecommand \citenamefont [1]{#1}%
\providecommand \href@noop [0]{\@secondoftwo}%
\providecommand \href [0]{\begingroup \@sanitize@url \@href}%
\providecommand \@href[1]{\@@startlink{#1}\@@href}%
\providecommand \@@href[1]{\endgroup#1\@@endlink}%
\providecommand \@sanitize@url [0]{\catcode `\\12\catcode `\$12\catcode
  `\&12\catcode `\#12\catcode `\^12\catcode `\_12\catcode `\%12\relax}%
\providecommand \@@startlink[1]{}%
\providecommand \@@endlink[0]{}%
\providecommand \url  [0]{\begingroup\@sanitize@url \@url }%
\providecommand \@url [1]{\endgroup\@href {#1}{\urlprefix }}%
\providecommand \urlprefix  [0]{URL }%
\providecommand \Eprint [0]{\href }%
\providecommand \doibase [0]{https://doi.org/}%
\providecommand \selectlanguage [0]{\@gobble}%
\providecommand \bibinfo  [0]{\@secondoftwo}%
\providecommand \bibfield  [0]{\@secondoftwo}%
\providecommand \translation [1]{[#1]}%
\providecommand \BibitemOpen [0]{}%
\providecommand \bibitemStop [0]{}%
\providecommand \bibitemNoStop [0]{.\EOS\space}%
\providecommand \EOS [0]{\spacefactor3000\relax}%
\providecommand \BibitemShut  [1]{\csname bibitem#1\endcsname}%
\let\auto@bib@innerbib\@empty
\bibitem [{\citenamefont {Neugart}\ \emph {et~al.}(2017)\citenamefont
  {Neugart}, \citenamefont {Billowes}, \citenamefont {Bissell}, \citenamefont
  {Blaum}, \citenamefont {Cheal}, \citenamefont {Flanagan}, \citenamefont
  {Neyens}, \citenamefont {N{\"o}rtersh{\"a}user},\ and\ \citenamefont
  {Yordanov}}]{neugart2017collinear}%
  \BibitemOpen
  \bibfield  {author} {\bibinfo {author} {\bibfnamefont {R.}~\bibnamefont
  {Neugart}}, \bibinfo {author} {\bibfnamefont {J.}~\bibnamefont {Billowes}},
  \bibinfo {author} {\bibfnamefont {M.}~\bibnamefont {Bissell}}, \bibinfo
  {author} {\bibfnamefont {K.}~\bibnamefont {Blaum}}, \bibinfo {author}
  {\bibfnamefont {B.}~\bibnamefont {Cheal}}, \bibinfo {author} {\bibfnamefont
  {K.}~\bibnamefont {Flanagan}}, \bibinfo {author} {\bibfnamefont
  {G.}~\bibnamefont {Neyens}}, \bibinfo {author} {\bibfnamefont
  {W.}~\bibnamefont {N{\"o}rtersh{\"a}user}},\ and\ \bibinfo {author}
  {\bibfnamefont {D.}~\bibnamefont {Yordanov}},\ }\href@noop {} {\bibfield
  {journal} {\bibinfo  {journal} {Journal of Physics G: Nuclear and Particle
  Physics}\ }\textbf {\bibinfo {volume} {44}},\ \bibinfo {pages} {064002}
  (\bibinfo {year} {2017})}\BibitemShut {NoStop}%
\bibitem [{\citenamefont {Studer}\ \emph {et~al.}(2019)\citenamefont {Studer},
  \citenamefont {Heinitz}, \citenamefont {Heinke}, \citenamefont {Naubereit},
  \citenamefont {Dressler}, \citenamefont {Guerrero}, \citenamefont
  {K{\"o}ster}, \citenamefont {Schumann},\ and\ \citenamefont
  {Wendt}}]{studer2019atomic}%
  \BibitemOpen
  \bibfield  {author} {\bibinfo {author} {\bibfnamefont {D.}~\bibnamefont
  {Studer}}, \bibinfo {author} {\bibfnamefont {S.}~\bibnamefont {Heinitz}},
  \bibinfo {author} {\bibfnamefont {R.}~\bibnamefont {Heinke}}, \bibinfo
  {author} {\bibfnamefont {P.}~\bibnamefont {Naubereit}}, \bibinfo {author}
  {\bibfnamefont {R.}~\bibnamefont {Dressler}}, \bibinfo {author}
  {\bibfnamefont {C.}~\bibnamefont {Guerrero}}, \bibinfo {author}
  {\bibfnamefont {U.}~\bibnamefont {K{\"o}ster}}, \bibinfo {author}
  {\bibfnamefont {D.}~\bibnamefont {Schumann}},\ and\ \bibinfo {author}
  {\bibfnamefont {K.}~\bibnamefont {Wendt}},\ }\href@noop {} {\bibfield
  {journal} {\bibinfo  {journal} {Physical Review A}\ }\textbf {\bibinfo
  {volume} {99}},\ \bibinfo {pages} {062513} (\bibinfo {year}
  {2019})}\BibitemShut {NoStop}%
\bibitem [{\citenamefont {Raab}\ \emph {et~al.}(1987)\citenamefont {Raab},
  \citenamefont {Prentiss}, \citenamefont {Cable}, \citenamefont {Chu},\ and\
  \citenamefont {Pritchard}}]{Raab1987}%
  \BibitemOpen
  \bibfield  {author} {\bibinfo {author} {\bibfnamefont {E.}~\bibnamefont
  {Raab}}, \bibinfo {author} {\bibfnamefont {M.}~\bibnamefont {Prentiss}},
  \bibinfo {author} {\bibfnamefont {A.}~\bibnamefont {Cable}}, \bibinfo
  {author} {\bibfnamefont {S.}~\bibnamefont {Chu}},\ and\ \bibinfo {author}
  {\bibfnamefont {D.}~\bibnamefont {Pritchard}},\ }\href@noop {} {\bibfield
  {journal} {\bibinfo  {journal} {Physical Review Letters}\ }\textbf {\bibinfo
  {volume} {59}},\ \bibinfo {pages} {2631} (\bibinfo {year}
  {1987})}\BibitemShut {NoStop}%
\bibitem [{\citenamefont {Harvey}\ and\ \citenamefont
  {Murray}(2008)}]{harvey2008cold}%
  \BibitemOpen
  \bibfield  {author} {\bibinfo {author} {\bibfnamefont {M.}~\bibnamefont
  {Harvey}}\ and\ \bibinfo {author} {\bibfnamefont {A.~J.}\ \bibnamefont
  {Murray}},\ }\href@noop {} {\bibfield  {journal} {\bibinfo  {journal}
  {Physical review letters}\ }\textbf {\bibinfo {volume} {101}},\ \bibinfo
  {pages} {173201} (\bibinfo {year} {2008})}\BibitemShut {NoStop}%
\bibitem [{\citenamefont {Corti{\~n}as}\ \emph {et~al.}(2020)\citenamefont
  {Corti{\~n}as}, \citenamefont {Favier}, \citenamefont {Ravon}, \citenamefont
  {M{\'e}haignerie}, \citenamefont {Machu}, \citenamefont {Raimond},
  \citenamefont {Sayrin},\ and\ \citenamefont {Brune}}]{cortinas2020laser}%
  \BibitemOpen
  \bibfield  {author} {\bibinfo {author} {\bibfnamefont {R.}~\bibnamefont
  {Corti{\~n}as}}, \bibinfo {author} {\bibfnamefont {M.}~\bibnamefont
  {Favier}}, \bibinfo {author} {\bibfnamefont {B.}~\bibnamefont {Ravon}},
  \bibinfo {author} {\bibfnamefont {P.}~\bibnamefont {M{\'e}haignerie}},
  \bibinfo {author} {\bibfnamefont {Y.}~\bibnamefont {Machu}}, \bibinfo
  {author} {\bibfnamefont {J.}~\bibnamefont {Raimond}}, \bibinfo {author}
  {\bibfnamefont {C.}~\bibnamefont {Sayrin}},\ and\ \bibinfo {author}
  {\bibfnamefont {M.}~\bibnamefont {Brune}},\ }\href@noop {} {\bibfield
  {journal} {\bibinfo  {journal} {Physical Review Letters}\ }\textbf {\bibinfo
  {volume} {124}},\ \bibinfo {pages} {123201} (\bibinfo {year}
  {2020})}\BibitemShut {NoStop}%
\bibitem [{\citenamefont {Urvoy}\ \emph {et~al.}(2019)\citenamefont {Urvoy},
  \citenamefont {Vendeiro}, \citenamefont {Ramette}, \citenamefont
  {Adiyatullin},\ and\ \citenamefont {Vuleti{\'c}}}]{urvoy2019direct}%
  \BibitemOpen
  \bibfield  {author} {\bibinfo {author} {\bibfnamefont {A.}~\bibnamefont
  {Urvoy}}, \bibinfo {author} {\bibfnamefont {Z.}~\bibnamefont {Vendeiro}},
  \bibinfo {author} {\bibfnamefont {J.}~\bibnamefont {Ramette}}, \bibinfo
  {author} {\bibfnamefont {A.}~\bibnamefont {Adiyatullin}},\ and\ \bibinfo
  {author} {\bibfnamefont {V.}~\bibnamefont {Vuleti{\'c}}},\ }\href@noop {}
  {\bibfield  {journal} {\bibinfo  {journal} {Physical review letters}\
  }\textbf {\bibinfo {volume} {122}},\ \bibinfo {pages} {203202} (\bibinfo
  {year} {2019})}\BibitemShut {NoStop}%
\bibitem [{\citenamefont {Hertel}\ and\ \citenamefont
  {Stoll}(1978)}]{HertelStoll}%
  \BibitemOpen
  \bibfield  {author} {\bibinfo {author} {\bibfnamefont {I.}~\bibnamefont
  {Hertel}}\ and\ \bibinfo {author} {\bibfnamefont {W.}~\bibnamefont {Stoll}},\
  }\href@noop {} {\bibfield  {journal} {\bibinfo  {journal} {Advances in Atomic
  and Molecular Physics}\ }\textbf {\bibinfo {volume} {13}},\ \bibinfo {pages}
  {113} (\bibinfo {year} {1978})}\BibitemShut {NoStop}%
\bibitem [{\citenamefont {Farrell}\ \emph {et~al.}(1988)\citenamefont
  {Farrell}, \citenamefont {MacGillivray},\ and\ \citenamefont
  {Standage}}]{farrell1988quantum}%
  \BibitemOpen
  \bibfield  {author} {\bibinfo {author} {\bibfnamefont {P.}~\bibnamefont
  {Farrell}}, \bibinfo {author} {\bibfnamefont {W.}~\bibnamefont
  {MacGillivray}},\ and\ \bibinfo {author} {\bibfnamefont {M.}~\bibnamefont
  {Standage}},\ }\href@noop {} {\bibfield  {journal} {\bibinfo  {journal}
  {Physical Review A}\ }\textbf {\bibinfo {volume} {37}},\ \bibinfo {pages}
  {4240} (\bibinfo {year} {1988})}\BibitemShut {NoStop}%
\bibitem [{\citenamefont {MacGillivray}\ and\ \citenamefont
  {Standage}(1988)}]{Macgillivray_Standage1988}%
  \BibitemOpen
  \bibfield  {author} {\bibinfo {author} {\bibfnamefont {W.}~\bibnamefont
  {MacGillivray}}\ and\ \bibinfo {author} {\bibfnamefont {M.}~\bibnamefont
  {Standage}},\ }\href@noop {} {\bibfield  {journal} {\bibinfo  {journal}
  {Physicals Reports}\ }\textbf {\bibinfo {volume} {168}},\ \bibinfo {pages}
  {1} (\bibinfo {year} {1988})}\BibitemShut {NoStop}%
\bibitem [{\citenamefont {Murray}\ \emph {et~al.}(1989)\citenamefont {Murray},
  \citenamefont {Webb}, \citenamefont {MacGillvray},\ and\ \citenamefont
  {Standage}}]{murraystepwisePRL}%
  \BibitemOpen
  \bibfield  {author} {\bibinfo {author} {\bibfnamefont {A.}~\bibnamefont
  {Murray}}, \bibinfo {author} {\bibfnamefont {C.}~\bibnamefont {Webb}},
  \bibinfo {author} {\bibfnamefont {W.}~\bibnamefont {MacGillvray}},\ and\
  \bibinfo {author} {\bibfnamefont {M.}~\bibnamefont {Standage}},\ }\href@noop
  {} {\bibfield  {journal} {\bibinfo  {journal} {Physical Review Letters}\
  }\textbf {\bibinfo {volume} {62}},\ \bibinfo {pages} {411} (\bibinfo {year}
  {1989})}\BibitemShut {NoStop}%
\bibitem [{\citenamefont {Murray}\ \emph {et~al.}(1990)\citenamefont {Murray},
  \citenamefont {MacGillivray},\ and\ \citenamefont
  {Standage}}]{Stepwise_1990}%
  \BibitemOpen
  \bibfield  {author} {\bibinfo {author} {\bibfnamefont {A.}~\bibnamefont
  {Murray}}, \bibinfo {author} {\bibfnamefont {W.}~\bibnamefont
  {MacGillivray}},\ and\ \bibinfo {author} {\bibfnamefont {M.}~\bibnamefont
  {Standage}},\ }\href@noop {} {\bibfield  {journal} {\bibinfo  {journal}
  {Journal of Physics B: Atomic, Molecular and Optical Physics}\ }\textbf
  {\bibinfo {volume} {23}},\ \bibinfo {pages} {3373} (\bibinfo {year}
  {1990})}\BibitemShut {NoStop}%
\bibitem [{\citenamefont {Farrell}\ \emph {et~al.}(1991)\citenamefont
  {Farrell}, \citenamefont {MacGillivray},\ and\ \citenamefont
  {Standage}}]{farrell1991quantum}%
  \BibitemOpen
  \bibfield  {author} {\bibinfo {author} {\bibfnamefont {P.}~\bibnamefont
  {Farrell}}, \bibinfo {author} {\bibfnamefont {W.}~\bibnamefont
  {MacGillivray}},\ and\ \bibinfo {author} {\bibfnamefont {M.}~\bibnamefont
  {Standage}},\ }\href@noop {} {\bibfield  {journal} {\bibinfo  {journal}
  {Physical Review A}\ }\textbf {\bibinfo {volume} {44}},\ \bibinfo {pages}
  {1828} (\bibinfo {year} {1991})}\BibitemShut {NoStop}%
\bibitem [{\citenamefont {Murray}\ \emph
  {et~al.}(1991{\natexlab{a}})\citenamefont {Murray}, \citenamefont
  {MacGillvray},\ and\ \citenamefont {Standage}}]{murraystepwiseJModOpt}%
  \BibitemOpen
  \bibfield  {author} {\bibinfo {author} {\bibfnamefont {A.}~\bibnamefont
  {Murray}}, \bibinfo {author} {\bibfnamefont {W.}~\bibnamefont
  {MacGillvray}},\ and\ \bibinfo {author} {\bibfnamefont {M.}~\bibnamefont
  {Standage}},\ }\href@noop {} {\bibfield  {journal} {\bibinfo  {journal}
  {Journal of Modern Optics}\ }\textbf {\bibinfo {volume} {38}},\ \bibinfo
  {pages} {961} (\bibinfo {year} {1991}{\natexlab{a}})}\BibitemShut {NoStop}%
\bibitem [{\citenamefont {Murray}\ \emph {et~al.}(1992)\citenamefont {Murray},
  \citenamefont {Pascual}, \citenamefont {MacGillivray},\ and\ \citenamefont
  {Standage}}]{Stepwise_1992_expt}%
  \BibitemOpen
  \bibfield  {author} {\bibinfo {author} {\bibfnamefont {A.}~\bibnamefont
  {Murray}}, \bibinfo {author} {\bibfnamefont {R.}~\bibnamefont {Pascual}},
  \bibinfo {author} {\bibfnamefont {W.}~\bibnamefont {MacGillivray}},\ and\
  \bibinfo {author} {\bibfnamefont {M.}~\bibnamefont {Standage}},\ }\href@noop
  {} {\bibfield  {journal} {\bibinfo  {journal} {Journal of Physics B: Atomic,
  Molecular and Optical Physics}\ }\textbf {\bibinfo {volume} {25}},\ \bibinfo
  {pages} {1915} (\bibinfo {year} {1992})}\BibitemShut {NoStop}%
\bibitem [{\citenamefont {Murray}\ \emph
  {et~al.}(1991{\natexlab{b}})\citenamefont {Murray}, \citenamefont
  {MacGillivray},\ and\ \citenamefont {Standage}}]{murray_radiation_trapping}%
  \BibitemOpen
  \bibfield  {author} {\bibinfo {author} {\bibfnamefont {A.}~\bibnamefont
  {Murray}}, \bibinfo {author} {\bibfnamefont {W.}~\bibnamefont
  {MacGillivray}},\ and\ \bibinfo {author} {\bibfnamefont {M.}~\bibnamefont
  {Standage}},\ }\href@noop {} {\bibfield  {journal} {\bibinfo  {journal}
  {Physical Review A}\ }\textbf {\bibinfo {volume} {44}},\ \bibinfo {pages}
  {3162} (\bibinfo {year} {1991}{\natexlab{b}})}\BibitemShut {NoStop}%
\bibitem [{\citenamefont {Masters}\ \emph {et~al.}(1996)\citenamefont
  {Masters}, \citenamefont {Murray}, \citenamefont {Pascual},\ and\
  \citenamefont {Standage}}]{masters1996multipole}%
  \BibitemOpen
  \bibfield  {author} {\bibinfo {author} {\bibfnamefont {A.}~\bibnamefont
  {Masters}}, \bibinfo {author} {\bibfnamefont {A.}~\bibnamefont {Murray}},
  \bibinfo {author} {\bibfnamefont {R.}~\bibnamefont {Pascual}},\ and\ \bibinfo
  {author} {\bibfnamefont {M.}~\bibnamefont {Standage}},\ }\href@noop {}
  {\bibfield  {journal} {\bibinfo  {journal} {Physical Review A}\ }\textbf
  {\bibinfo {volume} {53}},\ \bibinfo {pages} {3884} (\bibinfo {year}
  {1996})}\BibitemShut {NoStop}%
\bibitem [{\citenamefont {Murray}\ and\ \citenamefont
  {Cvejanovic}(2003)}]{murray2003low}%
  \BibitemOpen
  \bibfield  {author} {\bibinfo {author} {\bibfnamefont {A.~J.}\ \bibnamefont
  {Murray}}\ and\ \bibinfo {author} {\bibfnamefont {D.}~\bibnamefont
  {Cvejanovic}},\ }\href@noop {} {\bibfield  {journal} {\bibinfo  {journal}
  {Journal of Physics B: Atomic, Molecular and Optical Physics}\ }\textbf
  {\bibinfo {volume} {36}},\ \bibinfo {pages} {4889} (\bibinfo {year}
  {2003})}\BibitemShut {NoStop}%
\bibitem [{\citenamefont {Hussey}\ \emph {et~al.}(2007)\citenamefont {Hussey},
  \citenamefont {Murray}, \citenamefont {MacGillivray},\ and\ \citenamefont
  {King}}]{superelasticBfieldPRL}%
  \BibitemOpen
  \bibfield  {author} {\bibinfo {author} {\bibfnamefont {M.}~\bibnamefont
  {Hussey}}, \bibinfo {author} {\bibfnamefont {A.~J.}\ \bibnamefont {Murray}},
  \bibinfo {author} {\bibfnamefont {W.}~\bibnamefont {MacGillivray}},\ and\
  \bibinfo {author} {\bibfnamefont {G.~C.}\ \bibnamefont {King}},\ }\href@noop
  {} {\bibfield  {journal} {\bibinfo  {journal} {Physical Review Letters}\
  }\textbf {\bibinfo {volume} {99}},\ \bibinfo {pages} {133202} (\bibinfo
  {year} {2007})}\BibitemShut {NoStop}%
\bibitem [{\citenamefont {Murray}\ \emph {et~al.}(2008)\citenamefont {Murray},
  \citenamefont {MacGillivray},\ and\ \citenamefont
  {Hussey}}]{murray2008theoretical}%
  \BibitemOpen
  \bibfield  {author} {\bibinfo {author} {\bibfnamefont {A.~J.}\ \bibnamefont
  {Murray}}, \bibinfo {author} {\bibfnamefont {W.}~\bibnamefont
  {MacGillivray}},\ and\ \bibinfo {author} {\bibfnamefont {M.}~\bibnamefont
  {Hussey}},\ }\href@noop {} {\bibfield  {journal} {\bibinfo  {journal}
  {Physical Review A}\ }\textbf {\bibinfo {volume} {77}},\ \bibinfo {pages}
  {013409} (\bibinfo {year} {2008})}\BibitemShut {NoStop}%
\bibitem [{\citenamefont {Hussey}\ \emph {et~al.}(2008)\citenamefont {Hussey},
  \citenamefont {Murray}, \citenamefont {MacGillivray},\ and\ \citenamefont
  {King}}]{superelasticBfieldJPB}%
  \BibitemOpen
  \bibfield  {author} {\bibinfo {author} {\bibfnamefont {M.}~\bibnamefont
  {Hussey}}, \bibinfo {author} {\bibfnamefont {A.}~\bibnamefont {Murray}},
  \bibinfo {author} {\bibfnamefont {W.}~\bibnamefont {MacGillivray}},\ and\
  \bibinfo {author} {\bibfnamefont {G.}~\bibnamefont {King}},\ }\href@noop {}
  {\bibfield  {journal} {\bibinfo  {journal} {Journal of Physics B: Atomic,
  Molecular and Optical Physics}\ }\textbf {\bibinfo {volume} {41}},\ \bibinfo
  {pages} {055202} (\bibinfo {year} {2008})}\BibitemShut {NoStop}%
\bibitem [{\citenamefont {Nixon}\ and\ \citenamefont
  {Murray}(2011)}]{e2eMgPRL1}%
  \BibitemOpen
  \bibfield  {author} {\bibinfo {author} {\bibfnamefont {K.~L.}\ \bibnamefont
  {Nixon}}\ and\ \bibinfo {author} {\bibfnamefont {A.~J.}\ \bibnamefont
  {Murray}},\ }\href@noop {} {\bibfield  {journal} {\bibinfo  {journal}
  {Physical Review Letters}\ }\textbf {\bibinfo {volume} {106}},\ \bibinfo
  {pages} {123201} (\bibinfo {year} {2011})}\BibitemShut {NoStop}%
\bibitem [{\citenamefont {Nixon}\ and\ \citenamefont
  {Murray}(2014)}]{e2eMgPRL2}%
  \BibitemOpen
  \bibfield  {author} {\bibinfo {author} {\bibfnamefont {K.~L.}\ \bibnamefont
  {Nixon}}\ and\ \bibinfo {author} {\bibfnamefont {A.~J.}\ \bibnamefont
  {Murray}},\ }\href@noop {} {\bibfield  {journal} {\bibinfo  {journal}
  {Physical Review Letters}\ }\textbf {\bibinfo {volume} {112}},\ \bibinfo
  {pages} {023202} (\bibinfo {year} {2014})}\BibitemShut {NoStop}%
\bibitem [{\citenamefont {Rudolph}\ \emph {et~al.}(2020)\citenamefont
  {Rudolph}, \citenamefont {Wilkason}, \citenamefont {Nantel}, \citenamefont
  {Swan}, \citenamefont {Holland}, \citenamefont {Jiang}, \citenamefont
  {Garber}, \citenamefont {Carman}, \citenamefont {Hogan} \emph
  {et~al.}}]{rudolph2020large}%
  \BibitemOpen
  \bibfield  {author} {\bibinfo {author} {\bibfnamefont {J.}~\bibnamefont
  {Rudolph}}, \bibinfo {author} {\bibfnamefont {T.}~\bibnamefont {Wilkason}},
  \bibinfo {author} {\bibfnamefont {M.}~\bibnamefont {Nantel}}, \bibinfo
  {author} {\bibfnamefont {H.}~\bibnamefont {Swan}}, \bibinfo {author}
  {\bibfnamefont {C.~M.}\ \bibnamefont {Holland}}, \bibinfo {author}
  {\bibfnamefont {Y.}~\bibnamefont {Jiang}}, \bibinfo {author} {\bibfnamefont
  {B.~E.}\ \bibnamefont {Garber}}, \bibinfo {author} {\bibfnamefont {S.~P.}\
  \bibnamefont {Carman}}, \bibinfo {author} {\bibfnamefont {J.~M.}\
  \bibnamefont {Hogan}}, \emph {et~al.},\ }\href@noop {} {\bibfield  {journal}
  {\bibinfo  {journal} {Physical review letters}\ }\textbf {\bibinfo {volume}
  {124}},\ \bibinfo {pages} {083604} (\bibinfo {year} {2020})}\BibitemShut
  {NoStop}%
\bibitem [{\citenamefont {McClelland}\ and\ \citenamefont
  {Kelley}(1985)}]{mcClelland}%
  \BibitemOpen
  \bibfield  {author} {\bibinfo {author} {\bibfnamefont {J.}~\bibnamefont
  {McClelland}}\ and\ \bibinfo {author} {\bibfnamefont {M.}~\bibnamefont
  {Kelley}},\ }\href@noop {} {\bibfield  {journal} {\bibinfo  {journal}
  {Physical Review A}\ }\textbf {\bibinfo {volume} {31}},\ \bibinfo {pages}
  {3704} (\bibinfo {year} {1985})}\BibitemShut {NoStop}%
\bibitem [{\citenamefont {Robertson}\ \emph {et~al.}(2021)\citenamefont
  {Robertson}, \citenamefont {{\v{S}}ibali{\'c}}, \citenamefont {Potvliege},\
  and\ \citenamefont {Jones}}]{robertson2021arc}%
  \BibitemOpen
  \bibfield  {author} {\bibinfo {author} {\bibfnamefont {E.~J.}\ \bibnamefont
  {Robertson}}, \bibinfo {author} {\bibfnamefont {N.}~\bibnamefont
  {{\v{S}}ibali{\'c}}}, \bibinfo {author} {\bibfnamefont {R.~M.}\ \bibnamefont
  {Potvliege}},\ and\ \bibinfo {author} {\bibfnamefont {M.~P.}\ \bibnamefont
  {Jones}},\ }\href@noop {} {\bibfield  {journal} {\bibinfo  {journal}
  {Computer Physics Communications}\ }\textbf {\bibinfo {volume} {261}},\
  \bibinfo {pages} {107814} (\bibinfo {year} {2021})}\BibitemShut {NoStop}%
\bibitem [{\citenamefont {Stenholm}(1986)}]{stenholm1986semiclassical}%
  \BibitemOpen
  \bibfield  {author} {\bibinfo {author} {\bibfnamefont {S.}~\bibnamefont
  {Stenholm}},\ }\href@noop {} {\bibfield  {journal} {\bibinfo  {journal}
  {Reviews of modern physics}\ }\textbf {\bibinfo {volume} {58}},\ \bibinfo
  {pages} {699} (\bibinfo {year} {1986})}\BibitemShut {NoStop}%
\bibitem [{\citenamefont
  {Milo{\v{s}}evi{\'c}}(2017)}]{milovsevic2017semiclassical}%
  \BibitemOpen
  \bibfield  {author} {\bibinfo {author} {\bibfnamefont {D.}~\bibnamefont
  {Milo{\v{s}}evi{\'c}}},\ }\href@noop {} {\bibfield  {journal} {\bibinfo
  {journal} {Physical Review A}\ }\textbf {\bibinfo {volume} {96}},\ \bibinfo
  {pages} {023413} (\bibinfo {year} {2017})}\BibitemShut {NoStop}%
\bibitem [{\citenamefont {Eckel}\ \emph {et~al.}(2022)\citenamefont {Eckel},
  \citenamefont {Barker}, \citenamefont {Norrgard},\ and\ \citenamefont
  {Scherschligt}}]{eckel2022pylcp}%
  \BibitemOpen
  \bibfield  {author} {\bibinfo {author} {\bibfnamefont {S.}~\bibnamefont
  {Eckel}}, \bibinfo {author} {\bibfnamefont {D.~S.}\ \bibnamefont {Barker}},
  \bibinfo {author} {\bibfnamefont {E.~B.}\ \bibnamefont {Norrgard}},\ and\
  \bibinfo {author} {\bibfnamefont {J.}~\bibnamefont {Scherschligt}},\
  }\href@noop {} {\bibfield  {journal} {\bibinfo  {journal} {Computer Physics
  Communications}\ }\textbf {\bibinfo {volume} {270}},\ \bibinfo {pages}
  {108166} (\bibinfo {year} {2022})}\BibitemShut {NoStop}%
\bibitem [{\citenamefont {Patel}(2021{\natexlab{a}})}]{LASEDreadthedocs}%
  \BibitemOpen
  \bibfield  {author} {\bibinfo {author} {\bibfnamefont {M.}~\bibnamefont
  {Patel}},\ }\href@noop {} {\bibinfo {title} {\text{LASED} documentation}}
  (\bibinfo {year} {2021}{\natexlab{a}}),\ \bibinfo {note}
  {\url{https://lased.readthedocs.io/en/latest/}}\BibitemShut {NoStop}%
\bibitem [{\citenamefont {Loudon}(2000)}]{loudon2000quantum}%
  \BibitemOpen
  \bibfield  {author} {\bibinfo {author} {\bibfnamefont {R.}~\bibnamefont
  {Loudon}},\ }\href@noop {} {\emph {\bibinfo {title} {The quantum theory of
  light}}}\ (\bibinfo  {publisher} {OUP Oxford},\ \bibinfo {year}
  {2000})\BibitemShut {NoStop}%
\bibitem [{\citenamefont {Ackerhalt}\ and\ \citenamefont
  {Eberly}(1974)}]{ackerhalt1974quantum}%
  \BibitemOpen
  \bibfield  {author} {\bibinfo {author} {\bibfnamefont {J.}~\bibnamefont
  {Ackerhalt}}\ and\ \bibinfo {author} {\bibfnamefont {J.}~\bibnamefont
  {Eberly}},\ }\href@noop {} {\bibfield  {journal} {\bibinfo  {journal}
  {Physical Review D}\ }\textbf {\bibinfo {volume} {10}},\ \bibinfo {pages}
  {3350} (\bibinfo {year} {1974})}\BibitemShut {NoStop}%
\bibitem [{\citenamefont {Kubo}(1963)}]{kubo1963stochastic}%
  \BibitemOpen
  \bibfield  {author} {\bibinfo {author} {\bibfnamefont {R.}~\bibnamefont
  {Kubo}},\ }\href@noop {} {\bibfield  {journal} {\bibinfo  {journal} {Journal
  of Mathematical Physics}\ }\textbf {\bibinfo {volume} {4}},\ \bibinfo {pages}
  {174} (\bibinfo {year} {1963})}\BibitemShut {NoStop}%
\bibitem [{\citenamefont {Whitley}\ and\ \citenamefont
  {Stroud~Jr}(1976)}]{whitley1976double}%
  \BibitemOpen
  \bibfield  {author} {\bibinfo {author} {\bibfnamefont {R.~M.}\ \bibnamefont
  {Whitley}}\ and\ \bibinfo {author} {\bibfnamefont {C.}~\bibnamefont
  {Stroud~Jr}},\ }\href@noop {} {\bibfield  {journal} {\bibinfo  {journal}
  {Physical Review A}\ }\textbf {\bibinfo {volume} {14}},\ \bibinfo {pages}
  {1498} (\bibinfo {year} {1976})}\BibitemShut {NoStop}%
\bibitem [{\citenamefont {Farrell}\ and\ \citenamefont
  {MacGillivray}(1995)}]{farrell1995consistency}%
  \BibitemOpen
  \bibfield  {author} {\bibinfo {author} {\bibfnamefont {P.}~\bibnamefont
  {Farrell}}\ and\ \bibinfo {author} {\bibfnamefont {W.}~\bibnamefont
  {MacGillivray}},\ }\href@noop {} {\bibfield  {journal} {\bibinfo  {journal}
  {Journal of physics A: Math. Gen.}\ }\textbf {\bibinfo {volume} {28}},\
  \bibinfo {pages} {209} (\bibinfo {year} {1995})}\BibitemShut {NoStop}%
\bibitem [{\citenamefont {Meurer}\ \emph {et~al.}(2017)\citenamefont {Meurer},
  \citenamefont {Smith}, \citenamefont {Paprocki}, \citenamefont
  {\v{C}ert\'{i}k}, \citenamefont {Kirpichev}, \citenamefont {Rocklin},
  \citenamefont {Kumar}, \citenamefont {Ivanov}, \citenamefont {Moore},
  \citenamefont {Singh}, \citenamefont {Rathnayake}, \citenamefont {Vig},
  \citenamefont {Granger}, \citenamefont {Muller}, \citenamefont {Bonazzi},
  \citenamefont {Gupta}, \citenamefont {Vats}, \citenamefont {Johansson},
  \citenamefont {Pedregosa}, \citenamefont {Curry}, \citenamefont {Terrel},
  \citenamefont {Rou\v{c}ka}, \citenamefont {Saboo}, \citenamefont {Fernando},
  \citenamefont {Kulal}, \citenamefont {Cimrman},\ and\ \citenamefont
  {Scopatz}}]{sympy2017}%
  \BibitemOpen
  \bibfield  {author} {\bibinfo {author} {\bibfnamefont {A.}~\bibnamefont
  {Meurer}}, \bibinfo {author} {\bibfnamefont {C.~P.}\ \bibnamefont {Smith}},
  \bibinfo {author} {\bibfnamefont {M.}~\bibnamefont {Paprocki}}, \bibinfo
  {author} {\bibfnamefont {O.}~\bibnamefont {\v{C}ert\'{i}k}}, \bibinfo
  {author} {\bibfnamefont {S.~B.}\ \bibnamefont {Kirpichev}}, \bibinfo {author}
  {\bibfnamefont {M.}~\bibnamefont {Rocklin}}, \bibinfo {author} {\bibfnamefont
  {A.}~\bibnamefont {Kumar}}, \bibinfo {author} {\bibfnamefont
  {S.}~\bibnamefont {Ivanov}}, \bibinfo {author} {\bibfnamefont {J.~K.}\
  \bibnamefont {Moore}}, \bibinfo {author} {\bibfnamefont {S.}~\bibnamefont
  {Singh}}, \bibinfo {author} {\bibfnamefont {T.}~\bibnamefont {Rathnayake}},
  \bibinfo {author} {\bibfnamefont {S.}~\bibnamefont {Vig}}, \bibinfo {author}
  {\bibfnamefont {B.~E.}\ \bibnamefont {Granger}}, \bibinfo {author}
  {\bibfnamefont {R.~P.}\ \bibnamefont {Muller}}, \bibinfo {author}
  {\bibfnamefont {F.}~\bibnamefont {Bonazzi}}, \bibinfo {author} {\bibfnamefont
  {H.}~\bibnamefont {Gupta}}, \bibinfo {author} {\bibfnamefont
  {S.}~\bibnamefont {Vats}}, \bibinfo {author} {\bibfnamefont {F.}~\bibnamefont
  {Johansson}}, \bibinfo {author} {\bibfnamefont {F.}~\bibnamefont
  {Pedregosa}}, \bibinfo {author} {\bibfnamefont {M.~J.}\ \bibnamefont
  {Curry}}, \bibinfo {author} {\bibfnamefont {A.~R.}\ \bibnamefont {Terrel}},
  \bibinfo {author} {\bibfnamefont {v.}~\bibnamefont {Rou\v{c}ka}}, \bibinfo
  {author} {\bibfnamefont {A.}~\bibnamefont {Saboo}}, \bibinfo {author}
  {\bibfnamefont {I.}~\bibnamefont {Fernando}}, \bibinfo {author}
  {\bibfnamefont {S.}~\bibnamefont {Kulal}}, \bibinfo {author} {\bibfnamefont
  {R.}~\bibnamefont {Cimrman}},\ and\ \bibinfo {author} {\bibfnamefont
  {A.}~\bibnamefont {Scopatz}},\ }\href {https://doi.org/10.7717/peerj-cs.103}
  {\bibfield  {journal} {\bibinfo  {journal} {PeerJ Computer Science}\ }\textbf
  {\bibinfo {volume} {3}},\ \bibinfo {pages} {e103} (\bibinfo {year}
  {2017})}\BibitemShut {NoStop}%
\bibitem [{\citenamefont {Harris}\ \emph {et~al.}(2020)\citenamefont {Harris},
  \citenamefont {Millman}, \citenamefont {van~der Walt}, \citenamefont
  {Gommers}, \citenamefont {Virtanen}, \citenamefont {Cournapeau},
  \citenamefont {Wieser}, \citenamefont {Taylor}, \citenamefont {Berg},
  \citenamefont {Smith}, \citenamefont {Kern}, \citenamefont {Picus},
  \citenamefont {Hoyer}, \citenamefont {van Kerkwijk}, \citenamefont {Brett},
  \citenamefont {Haldane}, \citenamefont {del R{\'{i}}o}, \citenamefont
  {Wiebe}, \citenamefont {Peterson}, \citenamefont {G{\'{e}}rard-Marchant},
  \citenamefont {Sheppard}, \citenamefont {Reddy}, \citenamefont {Weckesser},
  \citenamefont {Abbasi}, \citenamefont {Gohlke},\ and\ \citenamefont
  {Oliphant}}]{harris2020array}%
  \BibitemOpen
  \bibfield  {author} {\bibinfo {author} {\bibfnamefont {C.~R.}\ \bibnamefont
  {Harris}}, \bibinfo {author} {\bibfnamefont {K.~J.}\ \bibnamefont {Millman}},
  \bibinfo {author} {\bibfnamefont {S.~J.}\ \bibnamefont {van~der Walt}},
  \bibinfo {author} {\bibfnamefont {R.}~\bibnamefont {Gommers}}, \bibinfo
  {author} {\bibfnamefont {P.}~\bibnamefont {Virtanen}}, \bibinfo {author}
  {\bibfnamefont {D.}~\bibnamefont {Cournapeau}}, \bibinfo {author}
  {\bibfnamefont {E.}~\bibnamefont {Wieser}}, \bibinfo {author} {\bibfnamefont
  {J.}~\bibnamefont {Taylor}}, \bibinfo {author} {\bibfnamefont
  {S.}~\bibnamefont {Berg}}, \bibinfo {author} {\bibfnamefont {N.~J.}\
  \bibnamefont {Smith}}, \bibinfo {author} {\bibfnamefont {R.}~\bibnamefont
  {Kern}}, \bibinfo {author} {\bibfnamefont {M.}~\bibnamefont {Picus}},
  \bibinfo {author} {\bibfnamefont {S.}~\bibnamefont {Hoyer}}, \bibinfo
  {author} {\bibfnamefont {M.~H.}\ \bibnamefont {van Kerkwijk}}, \bibinfo
  {author} {\bibfnamefont {M.}~\bibnamefont {Brett}}, \bibinfo {author}
  {\bibfnamefont {A.}~\bibnamefont {Haldane}}, \bibinfo {author} {\bibfnamefont
  {J.~F.}\ \bibnamefont {del R{\'{i}}o}}, \bibinfo {author} {\bibfnamefont
  {M.}~\bibnamefont {Wiebe}}, \bibinfo {author} {\bibfnamefont
  {P.}~\bibnamefont {Peterson}}, \bibinfo {author} {\bibfnamefont
  {P.}~\bibnamefont {G{\'{e}}rard-Marchant}}, \bibinfo {author} {\bibfnamefont
  {K.}~\bibnamefont {Sheppard}}, \bibinfo {author} {\bibfnamefont
  {T.}~\bibnamefont {Reddy}}, \bibinfo {author} {\bibfnamefont
  {W.}~\bibnamefont {Weckesser}}, \bibinfo {author} {\bibfnamefont
  {H.}~\bibnamefont {Abbasi}}, \bibinfo {author} {\bibfnamefont
  {C.}~\bibnamefont {Gohlke}},\ and\ \bibinfo {author} {\bibfnamefont {T.~E.}\
  \bibnamefont {Oliphant}},\ }\href {https://doi.org/10.1038/s41586-020-2649-2}
  {\bibfield  {journal} {\bibinfo  {journal} {Nature}\ }\textbf {\bibinfo
  {volume} {585}},\ \bibinfo {pages} {357} (\bibinfo {year}
  {2020})}\BibitemShut {NoStop}%
\bibitem [{\citenamefont {Virtanen}\ \emph {et~al.}(2020)\citenamefont
  {Virtanen}, \citenamefont {Gommers}, \citenamefont {Oliphant}, \citenamefont
  {Haberland}, \citenamefont {Reddy}, \citenamefont {Cournapeau}, \citenamefont
  {Burovski}, \citenamefont {Peterson}, \citenamefont {Weckesser},
  \citenamefont {Bright}, \citenamefont {{van der Walt}}, \citenamefont
  {Brett}, \citenamefont {Wilson}, \citenamefont {Millman}, \citenamefont
  {Mayorov}, \citenamefont {Nelson}, \citenamefont {Jones}, \citenamefont
  {Kern}, \citenamefont {Larson}, \citenamefont {Carey}, \citenamefont {Polat},
  \citenamefont {Feng}, \citenamefont {Moore}, \citenamefont {{VanderPlas}},
  \citenamefont {Laxalde}, \citenamefont {Perktold}, \citenamefont {Cimrman},
  \citenamefont {Henriksen}, \citenamefont {Quintero}, \citenamefont {Harris},
  \citenamefont {Archibald}, \citenamefont {Ribeiro}, \citenamefont
  {Pedregosa}, \citenamefont {{van Mulbregt}},\ and\ \citenamefont {{SciPy 1.0
  Contributors}}}]{2020SciPy-NMeth}%
  \BibitemOpen
  \bibfield  {author} {\bibinfo {author} {\bibfnamefont {P.}~\bibnamefont
  {Virtanen}}, \bibinfo {author} {\bibfnamefont {R.}~\bibnamefont {Gommers}},
  \bibinfo {author} {\bibfnamefont {T.~E.}\ \bibnamefont {Oliphant}}, \bibinfo
  {author} {\bibfnamefont {M.}~\bibnamefont {Haberland}}, \bibinfo {author}
  {\bibfnamefont {T.}~\bibnamefont {Reddy}}, \bibinfo {author} {\bibfnamefont
  {D.}~\bibnamefont {Cournapeau}}, \bibinfo {author} {\bibfnamefont
  {E.}~\bibnamefont {Burovski}}, \bibinfo {author} {\bibfnamefont
  {P.}~\bibnamefont {Peterson}}, \bibinfo {author} {\bibfnamefont
  {W.}~\bibnamefont {Weckesser}}, \bibinfo {author} {\bibfnamefont
  {J.}~\bibnamefont {Bright}}, \bibinfo {author} {\bibfnamefont {S.~J.}\
  \bibnamefont {{van der Walt}}}, \bibinfo {author} {\bibfnamefont
  {M.}~\bibnamefont {Brett}}, \bibinfo {author} {\bibfnamefont
  {J.}~\bibnamefont {Wilson}}, \bibinfo {author} {\bibfnamefont {K.~J.}\
  \bibnamefont {Millman}}, \bibinfo {author} {\bibfnamefont {N.}~\bibnamefont
  {Mayorov}}, \bibinfo {author} {\bibfnamefont {A.~R.~J.}\ \bibnamefont
  {Nelson}}, \bibinfo {author} {\bibfnamefont {E.}~\bibnamefont {Jones}},
  \bibinfo {author} {\bibfnamefont {R.}~\bibnamefont {Kern}}, \bibinfo {author}
  {\bibfnamefont {E.}~\bibnamefont {Larson}}, \bibinfo {author} {\bibfnamefont
  {C.~J.}\ \bibnamefont {Carey}}, \bibinfo {author} {\bibfnamefont
  {{\.I}.}~\bibnamefont {Polat}}, \bibinfo {author} {\bibfnamefont
  {Y.}~\bibnamefont {Feng}}, \bibinfo {author} {\bibfnamefont {E.~W.}\
  \bibnamefont {Moore}}, \bibinfo {author} {\bibfnamefont {J.}~\bibnamefont
  {{VanderPlas}}}, \bibinfo {author} {\bibfnamefont {D.}~\bibnamefont
  {Laxalde}}, \bibinfo {author} {\bibfnamefont {J.}~\bibnamefont {Perktold}},
  \bibinfo {author} {\bibfnamefont {R.}~\bibnamefont {Cimrman}}, \bibinfo
  {author} {\bibfnamefont {I.}~\bibnamefont {Henriksen}}, \bibinfo {author}
  {\bibfnamefont {E.~A.}\ \bibnamefont {Quintero}}, \bibinfo {author}
  {\bibfnamefont {C.~R.}\ \bibnamefont {Harris}}, \bibinfo {author}
  {\bibfnamefont {A.~M.}\ \bibnamefont {Archibald}}, \bibinfo {author}
  {\bibfnamefont {A.~H.}\ \bibnamefont {Ribeiro}}, \bibinfo {author}
  {\bibfnamefont {F.}~\bibnamefont {Pedregosa}}, \bibinfo {author}
  {\bibfnamefont {P.}~\bibnamefont {{van Mulbregt}}},\ and\ \bibinfo {author}
  {\bibnamefont {{SciPy 1.0 Contributors}}},\ }\href
  {https://doi.org/10.1038/s41592-019-0686-2} {\bibfield  {journal} {\bibinfo
  {journal} {Nature Methods}\ }\textbf {\bibinfo {volume} {17}},\ \bibinfo
  {pages} {261} (\bibinfo {year} {2020})}\BibitemShut {NoStop}%
\bibitem [{\citenamefont {Welford}\ \emph {et~al.}(1991)\citenamefont
  {Welford}, \citenamefont {Rines},\ and\ \citenamefont
  {Dinerman}}]{welford1991efficient}%
  \BibitemOpen
  \bibfield  {author} {\bibinfo {author} {\bibfnamefont {D.}~\bibnamefont
  {Welford}}, \bibinfo {author} {\bibfnamefont {D.}~\bibnamefont {Rines}},\
  and\ \bibinfo {author} {\bibfnamefont {B.}~\bibnamefont {Dinerman}},\
  }\href@noop {} {\bibfield  {journal} {\bibinfo  {journal} {Optics letters}\
  }\textbf {\bibinfo {volume} {16}},\ \bibinfo {pages} {1850} (\bibinfo {year}
  {1991})}\BibitemShut {NoStop}%
\bibitem [{\citenamefont {Siegman}(1986)}]{siegman1986lasers}%
  \BibitemOpen
  \bibfield  {author} {\bibinfo {author} {\bibfnamefont {A.~E.}\ \bibnamefont
  {Siegman}},\ }\href@noop {} {\emph {\bibinfo {title} {Lasers}}}\ (\bibinfo
  {publisher} {University science books},\ \bibinfo {year} {1986})\BibitemShut
  {NoStop}%
\bibitem [{\citenamefont {Brink}\ \emph {et~al.}(1963)\citenamefont {Brink},
  \citenamefont {Satchler},\ and\ \citenamefont {Danos}}]{brink1963angular}%
  \BibitemOpen
  \bibfield  {author} {\bibinfo {author} {\bibfnamefont {D.}~\bibnamefont
  {Brink}}, \bibinfo {author} {\bibfnamefont {G.}~\bibnamefont {Satchler}},\
  and\ \bibinfo {author} {\bibfnamefont {M.}~\bibnamefont {Danos}},\
  }\href@noop {} {\bibfield  {journal} {\bibinfo  {journal} {Physics Today}\
  }\textbf {\bibinfo {volume} {16}},\ \bibinfo {pages} {80} (\bibinfo {year}
  {1963})}\BibitemShut {NoStop}%
\bibitem [{\citenamefont {Biedenharn}\ \emph {et~al.}(1981)\citenamefont
  {Biedenharn}, \citenamefont {Louck},\ and\ \citenamefont
  {Carruthers}}]{biedenharn1981angular}%
  \BibitemOpen
  \bibfield  {author} {\bibinfo {author} {\bibfnamefont {L.~C.}\ \bibnamefont
  {Biedenharn}}, \bibinfo {author} {\bibfnamefont {J.~D.}\ \bibnamefont
  {Louck}},\ and\ \bibinfo {author} {\bibfnamefont {P.~A.}\ \bibnamefont
  {Carruthers}},\ }\href@noop {} {\emph {\bibinfo {title} {Angular momentum in
  quantum physics: theory and application}}},\ Vol.~\bibinfo {volume} {8}\
  (\bibinfo  {publisher} {Addison-Wesley Reading, MA},\ \bibinfo {year}
  {1981})\BibitemShut {NoStop}%
\bibitem [{\citenamefont {Lurio}\ \emph {et~al.}(1964)\citenamefont {Lurio},
  \citenamefont {DeZafra},\ and\ \citenamefont {Goshen}}]{lurio1964lifetime}%
  \BibitemOpen
  \bibfield  {author} {\bibinfo {author} {\bibfnamefont {A.}~\bibnamefont
  {Lurio}}, \bibinfo {author} {\bibfnamefont {R.}~\bibnamefont {DeZafra}},\
  and\ \bibinfo {author} {\bibfnamefont {R.~J.}\ \bibnamefont {Goshen}},\
  }\href@noop {} {\bibfield  {journal} {\bibinfo  {journal} {Physical Review}\
  }\textbf {\bibinfo {volume} {134}},\ \bibinfo {pages} {A1198} (\bibinfo
  {year} {1964})}\BibitemShut {NoStop}%
\bibitem [{\citenamefont {Risberg}(1968)}]{risberg1968spectrum}%
  \BibitemOpen
  \bibfield  {author} {\bibinfo {author} {\bibfnamefont {G.}~\bibnamefont
  {Risberg}},\ }\href@noop {} {\bibfield  {journal} {\bibinfo  {journal} {Arkiv
  for Fysik}\ }\textbf {\bibinfo {volume} {37}},\ \bibinfo {pages} {231}
  (\bibinfo {year} {1968})}\BibitemShut {NoStop}%
\bibitem [{\citenamefont {Beijers}\ \emph {et~al.}(1987)\citenamefont
  {Beijers}, \citenamefont {Doornenbal}, \citenamefont {van Eck},\ and\
  \citenamefont {Heideman}}]{beijers1987polarisation}%
  \BibitemOpen
  \bibfield  {author} {\bibinfo {author} {\bibfnamefont {J.}~\bibnamefont
  {Beijers}}, \bibinfo {author} {\bibfnamefont {S.}~\bibnamefont {Doornenbal}},
  \bibinfo {author} {\bibfnamefont {J.}~\bibnamefont {van Eck}},\ and\ \bibinfo
  {author} {\bibfnamefont {H.}~\bibnamefont {Heideman}},\ }\href@noop {}
  {\bibfield  {journal} {\bibinfo  {journal} {Journal of Physics B: Atomic and
  Molecular Physics (1968-1987)}\ }\textbf {\bibinfo {volume} {20}},\ \bibinfo
  {pages} {6617} (\bibinfo {year} {1987})}\BibitemShut {NoStop}%
\bibitem [{\citenamefont {Bray}(2020)}]{IgorBrayprivate}%
  \BibitemOpen
  \bibfield  {author} {\bibinfo {author} {\bibfnamefont {I.}~\bibnamefont
  {Bray}},\ }\href@noop {} {\bibfield  {journal} {\bibinfo  {journal} {private
  communication}\ } (\bibinfo {year} {2020})}\BibitemShut {NoStop}%
\bibitem [{\citenamefont {Kramida}\ \emph {et~al.}(2021)\citenamefont
  {Kramida}, \citenamefont {{Yu.~Ralchenko}}, \citenamefont {Reader},\ and\
  \citenamefont {{and NIST ASD Team}}}]{NIST_ASD}%
  \BibitemOpen
  \bibfield  {author} {\bibinfo {author} {\bibfnamefont {A.}~\bibnamefont
  {Kramida}}, \bibinfo {author} {\bibnamefont {{Yu.~Ralchenko}}}, \bibinfo
  {author} {\bibfnamefont {J.}~\bibnamefont {Reader}},\ and\ \bibinfo {author}
  {\bibnamefont {{and NIST ASD Team}}},\ }\href@noop {} {}\bibinfo
  {howpublished} {{NIST Atomic Spectra Database (ver. 5.9), [Online].
  Available: {\tt{https://physics.nist.gov/asd}} [2022, March 28]. National
  Institute of Standards and Technology, Gaithersburg, MD.}} (\bibinfo {year}
  {2021})\BibitemShut {NoStop}%
\bibitem [{\citenamefont {Rafac}\ \emph {et~al.}(1999)\citenamefont {Rafac},
  \citenamefont {Tanner}, \citenamefont {Livingston},\ and\ \citenamefont
  {Berry}}]{rafac1999fast}%
  \BibitemOpen
  \bibfield  {author} {\bibinfo {author} {\bibfnamefont {R.~J.}\ \bibnamefont
  {Rafac}}, \bibinfo {author} {\bibfnamefont {C.~E.}\ \bibnamefont {Tanner}},
  \bibinfo {author} {\bibfnamefont {A.~E.}\ \bibnamefont {Livingston}},\ and\
  \bibinfo {author} {\bibfnamefont {H.~G.}\ \bibnamefont {Berry}},\ }\href@noop
  {} {\bibfield  {journal} {\bibinfo  {journal} {Physical Review A}\ }\textbf
  {\bibinfo {volume} {60}},\ \bibinfo {pages} {3648} (\bibinfo {year}
  {1999})}\BibitemShut {NoStop}%
\bibitem [{\citenamefont {Udem}\ \emph {et~al.}(2000)\citenamefont {Udem},
  \citenamefont {Reichert}, \citenamefont {H{\"a}nsch},\ and\ \citenamefont
  {Kourogi}}]{udem2000absolute}%
  \BibitemOpen
  \bibfield  {author} {\bibinfo {author} {\bibfnamefont {T.}~\bibnamefont
  {Udem}}, \bibinfo {author} {\bibfnamefont {J.}~\bibnamefont {Reichert}},
  \bibinfo {author} {\bibfnamefont {T.}~\bibnamefont {H{\"a}nsch}},\ and\
  \bibinfo {author} {\bibfnamefont {M.}~\bibnamefont {Kourogi}},\ }\href@noop
  {} {\bibfield  {journal} {\bibinfo  {journal} {Physical Review A}\ }\textbf
  {\bibinfo {volume} {62}},\ \bibinfo {pages} {031801} (\bibinfo {year}
  {2000})}\BibitemShut {NoStop}%
\bibitem [{\citenamefont {Arimondo}\ \emph {et~al.}(1977)\citenamefont
  {Arimondo}, \citenamefont {Inguscio},\ and\ \citenamefont
  {Violino}}]{arimondo1977experimental}%
  \BibitemOpen
  \bibfield  {author} {\bibinfo {author} {\bibfnamefont {E.}~\bibnamefont
  {Arimondo}}, \bibinfo {author} {\bibfnamefont {M.}~\bibnamefont {Inguscio}},\
  and\ \bibinfo {author} {\bibfnamefont {P.}~\bibnamefont {Violino}},\
  }\href@noop {} {\bibfield  {journal} {\bibinfo  {journal} {Reviews of Modern
  Physics}\ }\textbf {\bibinfo {volume} {49}},\ \bibinfo {pages} {31} (\bibinfo
  {year} {1977})}\BibitemShut {NoStop}%
\bibitem [{\citenamefont {Patel}(2021{\natexlab{b}})}]{LASEDgithub}%
  \BibitemOpen
  \bibfield  {author} {\bibinfo {author} {\bibfnamefont {M.}~\bibnamefont
  {Patel}},\ }\href@noop {} {\bibinfo {title} {\text{LASED} source code}}
  (\bibinfo {year} {2021}{\natexlab{b}}),\ \bibinfo {note}
  {\url{https://github.com/mvpmanish/LASED}}\BibitemShut {NoStop}%
\bibitem [{\citenamefont {{Python Software
  Foundation}}(2022)}]{python_download}%
  \BibitemOpen
  \bibfield  {author} {\bibinfo {author} {\bibnamefont {{Python Software
  Foundation}}},\ }\href@noop {} {\bibinfo {title} {Python downloads}}
  (\bibinfo {year} {2022}),\ \bibinfo {note}
  {\url{https://www.python.org/downloads/}}\BibitemShut {NoStop}%
\end{thebibliography}%
 
\end{document}